\documentclass[%
superscriptaddress,
amsmath,amssymb,
aps,
longbibliography,
11pt
]{revtex4-2}

\usepackage{amssymb}
\usepackage{amsthm}
\usepackage{xfrac}
\usepackage{graphicx}
\usepackage{amsmath}
\usepackage{amsfonts}
\usepackage{color}
\usepackage{centernot}
\usepackage{mathtools}
\usepackage{stmaryrd}

\usepackage[dvipsnames]{xcolor}

\newcommand{\newtext}[1]{{#1}}

\begin{document}
\title{Multi-time scale-invariance of turbulence in a shell model }
\author{Alexei A. Mailybaev} 
\affiliation{Instituto de Matem\'atica Pura e Aplicada -- IMPA, Rio de Janeiro, Brazil}
\email{alexei@impa.br}

\begin{abstract}
When time and velocities are dynamically rescaled relative to the instantaneous turnover time, the Sabra shell model acquires another (hidden) form of scaling symmetry. 
It has been previously shown that this symmetry is statistically restored in the inertial interval of developed turbulence, \newtext{thereby establishing a self-similarity property derived from first principles and replacing the broken $1/3$-scaling of the K41 theory. Multifractal intermittency follows from the restored hidden symmetry, in which the anomalous scaling exponents $\zeta_p$ are identified as Perron-Frobenius eigenvalues.
In this paper, we use the hypothesis of restored hidden symmetry to address the multi-time statistics of turbulent fluctuations.
The central result is the self-similarity rule stating that any observable that is time-scale homogeneous of degree $p$ is self-similar with the H\"older exponent $h = \zeta_p/p$. 
As a particular case, it yields the scaling laws for decorrelation times of fluctuations obtained previously within the phenomenological multifractal approach.
As further applications,} we formulate self-similarity rules for multi-time structure functions and multi-time Kolmogorov multipliers and verify them numerically. 
\end{abstract}

\maketitle

\section{Introduction} 

Developed hydrodynamic turbulence is characterized by an extremelly large range of active scales. 
In this range, a special role is given to the inertial interval, which contains scales that are much smaller than the scales at which forcing is applied, and much larger than the scales at which dissipation (viscosity) becomes significant.
Description of universal statistical properties of the velocity field in the inertial interval is one of the central problems in the theory of developed turbulence~\cite{frisch1999turbulence}. 

\newtext{Shell models of turbulence are toy models that share many of the fundamental properties of turbulence and therefore provide a convenient platform for testing and verifying new ideas and theories~\cite{biferale2003shell}. 
One of the most popular is the Sabra shell model, introduced in~\cite{l1998improved} based on its predecessors~\cite{gledzer1973system,ohkitani1989temporal}, which mimics the Navier--Stokes flow using a geometric sequence of spatial scales $\ell_n = 2^{-n}$ with integer shell numbers $n$. The velocity fluctuations at the scale $\ell_n$ are represented by a complex variable $u_n \in \mathbb{C}$, which is called the shell velocity. In the inertial interval, shell velocities satisfy the ideal equations 
	\begin{equation}
	\frac{du_n}{dt} = B_n[u], \quad
	B_n[u] = ik_n \left(2u_{n+2}u_{n+1}^*
	-\frac{u_{n+1}u_{n-1}^*}{2} 
	+\frac{u_{n-1}u_{n-2}}{4}\right),
	\label{eq2_3}
	\end{equation}
where the quadratic nonlinearity $B_n[u]$ with the wavenumber $k_n = 1/\ell_n = 2^n$ mimics the convective term of the Navier--Stokes system. 
The system (\ref{eq2_3}) is designed to conserve energy $E[u] = \sum_n |u_n|^2$ and has a family of scaling symmetries
    \begin{equation}
    t,\ u_n \ \ \mapsto \ \ 2^{1-h}t,\ 2^h u_{n+1}, 
    \label{eq2_4}
    \end{equation}
with an arbitrary H\"older exponent $h$.
Expressions (\ref{eq2_4}) define new solutions as $\tilde{u}_n(t) = 2^{h} u_{n+1}\big(t/2^{1-h}\big)$.

\subsection{K41 theory and intermittency}

The central hypothesis of the Kolmogorov (K41) theory~\cite{kolmogorov1941local,frisch1999turbulence} is that the turbulent statistics restores the scaling symmetry (\ref{eq2_4}) for $h = 1/3$. This exponent is uniquely selected by the condition that the energy is transported from large to small sales. One of the immediate consequence is the K41 scalings of time-averaged structure functions
$\langle |u_n|^p \rangle_t \propto \ell_n^{p/3}$. Numerical analysis does not support the K41 hypothesis suggesting instead the anomalous scaling laws 
	\begin{equation}
	\big\langle |u_n|^p \big\rangle_t \propto \ell_n^{\zeta_p},
	\label{eqSS_I}
	\end{equation}
where the exponents $\zeta_p$ depend nonlinearly on $p$~\cite{frisch1999turbulence}. This anomalous scaling underlines what is called the intermittency phenomenon at small scales, and implies that the K41 symmetry as well as other symmetries (\ref{eq2_4}) are all broken in the inertial interval. Further developments assumed weaker forms of self-similarity, referring to the universality of velocity ratios (Kolmogorov multipliers) $|u_n/u_{n-1}|$~\cite{kolmogorov1962refinement,benzi1993intermittency,eyink2003gibbsian} or the multifractal model of Parisi-Frisch~\cite{frisch1985singularity,frisch1999turbulence} inspired by Mandelbrot's ideas. 
Despite excellent agreement with numerical simulations and success in predicting new physical properties, these theories are phenomenological, i.e., they are not derived from the equations of motion.

\subsection{Hidden scaling symmetry} \label{secI_HS}

Let us first review the concept of hidden symmetry, the central property of an ideal system on which the results of this paper are based.
It was suggested that Kolmogorov-style self-similarity does exist, but with respect to a different symmetry of the ideal system \cite{mailybaev2021hidden}. 
This symmetry appears after the appropriate rescaling (projection) of the solutions $u_n(t)$. 
Namely, fixing a reference scale $\ell_m$ in the inertial interval, one defines the rescaled velocity $U_N^{(m)}(\tau)$ as
	\begin{equation}
	U_N^{(m)}(\tau) = \frac{u_{N+m}(t)}{\ell_mA_m(t)}, \quad 
	\tau = \int_0^t A_m(s) ds, \quad
	A_m(t) = \frac{1}{\ell_m} \sqrt{\sum_{j \ge 0} 8^{-j}|u_{m-j}(t)|^2}.
	\label{eqI_R}
	\end{equation}
Here $A_m(t)$ measures the turnover frequency at the scale $\ell_m$ and time $t$. Note that there is a certain freedom in defining $A_m(t)$ \cite{mailybaev2020hidden}, but for simplicity we stick to a specific form (\ref{eqI_R}) in this paper. The turnover frequency is used to define the rescaled speeds $U_N^{(m)}$ and the rescaled time $\tau$ relative to the local level of fluctuations. The new time $\tau$ runs fast when the fluctuations are strong and slowly otherwise. 
In terms of new variables (\ref{eqI_R}) the ideal system (\ref{eq2_3}) becomes
	\begin{equation}
	\frac{dU_N}{d\tau} = B_N[U] - U_N R[U], \quad
	R[U] = \sum_{j \ge 0} 8^{-j} 
	\Re \left( U_{-j}^*B_{-j}[U]\right),
	\label{eq3_3}
	\end{equation}
where $\Re$ denotes the real part and we dropped the superscript $(m)$. The new system loses symmetries (\ref{eq2_4}), but acquires a new one. 
It is called hidden symmetry and is expressed as
	\begin{equation}
	\tilde{U}_N(\tilde\tau) = \frac{U_{N+1}(\tau)}{\sqrt{|U_1(\tau)|^2+8^{-1}}}, \quad
	\tilde\tau = 2\int_0^{\tau} \sqrt{|U_1(s)|^2+8^{-1}}\, ds.
	\label{eqI_HS}
	\end{equation}
One can verify that the velocities $\tilde{U}_N(\tilde\tau)$ indeed satisfy the same system (\ref{eq3_3}). Unlike symmetries (\ref{eq2_4}) depending on the scaling exponent $h$, the hidden symmetry has no free parameter. This new symmetry corresponds to changing the reference scale $\ell_m \mapsto \ell_{m+1}$: relations (\ref{eqI_HS}) transform the rescaled field ${U}_N(\tau) = U_N^{(m)}(\tau)$ into $\tilde{U}_N(\tilde\tau) = U_N^{(m+1)}(\tilde\tau)$.

It is conjectured that the hidden symmetry is restored in the turbulent statistics of the inertial interval, thereby replacing the broken scaling of K41. Practically, this means that the statistics of the rescaled field $U_N^{(m)}(\tau)$ does not depend on the choice of the scale $\ell_m$. This hypothesis was shown to be in full agreement with numerical data~\cite{mailybaev2021hidden,mailybaev2020hidden}. 

Remarkably, the restored hidden symmetry yields scaling laws (\ref{eqSS_I}), where the exponents $\zeta_p$ are defined as the Perron--Frobenius eigenvalues~\cite{mailybaev2022shell,mailybaev2023hidden}. The derivation of this relation goes back to Kolmogorov's ideas from 1962 \cite{kolmogorov1962refinement,benzi1993intermittency,eyink2003gibbsian}. Namely, let us express the velocity moment as  
	\begin{equation}
	|u_n|^p = |u_0|^p \prod_{m = 1}^n w_m^p.
	\label{eqI_PM}
	\end{equation}
Here $w_m$ is a so-called Kolmogorov multiplier defined as the ratio
	\begin{equation}
	w_m = \left|\frac{u_m}{u_{m-1}}\right| = \left|\frac{U_0^{(m)}}{U_{-1}^{(m)}}\right|,
	\label{eqI_PMb}
	\end{equation}
where the last expression follows from Eq.~(\ref{eqI_R}). The hidden scale invariance means that the statistics of multipliers do not depend on $m$. Hence, one can interpret Eq.~(\ref{eqI_PM}) as a (generalized) multiplicative Markov process, to which the Perron--Frobenius theory naturally applies; see e.g.~\cite{touchette2009large}. 
To summarize, hidden symmetry provides the missing link between the phenomenological multifractal model and the original equations of motion. It is important to note that a similar hidden symmetry exists in Navier-Stokes turbulence~\cite{mailybaev2020hidden,mailybaev2022hidden}.

\subsection{Multi-time statistics} \label{subsec_MTS}

Structure functions $\big\langle |u_n|^p \big\rangle_t$ refer to the single-time statistics of velocity fluctuations.
However, the complete description of the turbulent steady state must be based on its multi-time statistical properties. 
For example, a two-time correlation function at scale $\ell_n$ can be defined as 
	\begin{equation}
	F_n^{p_1,p_2}(s) = \big\langle |u_n(t)|^{p_1} |u_n(t+s)|^{p_2}  \big\rangle_t,
	\label{eqI_MS1}
	\end{equation}
where $\langle \cdot \rangle_t$ denotes the $t$-time average at a fixed time lag $s$.
The multi-time statistics were investigated in several works, e.g.~\cite{l1997temporal,biferale1999multi,mitra2004varieties,ray2008universality,pandit2008dynamic,biferale2011multi,rithwik2017revisiting}, where the analysis was based on the phenomenological multifractal model. 
Multi-time observables in these theories take the form of integrated correlation functions, from which physical quantities like decorrelation times of fluctuations are expressed. In particular, the integrated correlation function (\ref{eqI_MS1}) is defined as 
	\begin{equation}
	C_n^{p_1,p_2,q} = \int_0^\infty F_n^{p_1,p_2}(s) s^{q-1}ds,
	\label{eqI_MS2}
	\end{equation}
with the corresponding scaling relation 
	\begin{equation}
	C_n^{p_1,p_2,q} \propto \ell_n^z, \quad z = q+\zeta_{p_1+p_2-q}.
	\label{eqI_MS3}
	\end{equation}
One of the contributions of the present paper is to derive this and similar scaling laws directly from hidden scale invariance, thereby providing a first principles explanation of multi-time scalings previously obtained on a phenomenological basis; see Section \ref{subsecTI}.

The scaling in  Eq.~(\ref{eqI_MS3}) follows as a particular case of a universal self-similarity rule, which we derive as the consequence of the restored hidden symmetry; see Section~\ref{sec_4}.
It states that any (multi-time and multi-scale) observable that is time-scale homogeneous of degree $p$ is self-similar with respect to spacetime scaling with H\"older exponent $h = \zeta_p/p$.
This self-similarity rule is derived by relating time-scale homogeneous observables to the hidden-symmetric ones and using the previously established Perron--Frobenius asymptotics.
To facilitate the analysis, which is considerably more technical than in the case of single-time observables, we developed a new operator formalism. 
This formalism provides a unified description of both symmetries and observables and expresses the universal self-similarity rule in a compact form.
As applications, we show how known and new scaling laws follow from this universal rule. 
In particular, we derive power-law relations for multi-time structure functions and explain the universality properties of multi-time Kolmogorov multipliers. 
These relationships are verified with high accuracy by numerical simulations.

}

The paper is organized as follows. 
Section~\ref{sec_2} introduces the model and describes its statistics in terms of symmetry operators and observables. 
Section~\ref{sec_HS} introduces the hidden symmetry.
Section~\ref{sec_4} states the general self-similarity principle for time-scale homogeneous observables, which is the main result of the paper. 
Section~\ref{sec_5} considers applications to multi-time structure functions and multi-time Kolmogorov multipliers. 
Section~\ref{sec_der} provides the derivation of the self-similarity relation. 
Section~\ref{sec_7} summarizes the results.

\section{Broken scaling symmetries}
\label{sec_2}

\subsection{Model} 

The viscous Sabra shell model~\cite{l1998improved} is formulated in dimensionless form as
	\begin{equation}
	\frac{du_n}{dt} = B_n[u]- \mathrm{Re}^{-1} k_n^2u_n, \quad n > 0,
	\label{eq2_1}
	\end{equation}
where $\mathrm{Re}$ is the Reynolds number.
The quadratic term $B_n[u]$ given by Eq.~(\ref{eq2_3}) imitates the convective terms of the Navier--Stokes system.
The equations of motion (\ref{eq2_1}) must be provided with boundary (forcing) conditions for the shells $n \le 0$, which we define as
	\begin{equation}
	u_0(t) \equiv 1, \quad u_n(t) \equiv 0, \quad n < 0. 
	\label{eq2_BC}
	\end{equation}

We consider the fully developed turbulent state corresponding to very large Reynolds numbers. Our study is focused on the so-called inertial interval of scales $\ell_n$, at which both the forcing and viscous terms can be neglected. These scales satisfy the condition 
	\begin{equation}
	\ell_0 \gg \ell_n \gg \eta, 
	\label{eq2_II}
	\end{equation}
where $\ell_0 = 1$ is the forcing scale and $\eta$ is the (Kolmogorov) viscous micro-scale. 
The latter is the scale at which the viscous term in Eq.~(\ref{eq2_1}) becomes comparable to the nonlinear term.
The viscous scale $\eta \to 0$ vanishes as $\mathrm{Re} \to \infty$.
In the inertial range, the dynamics obey ideal Eq.~(\ref{eq2_3}), which follows from Eq.~(\ref{eq2_1}) after neglecting the viscous term. 

\subsection{Symmetry operators}

Ideal system~(\ref{eq2_3}) is invariant with respect to space-time scaling transformations (\ref{eq2_4}).
It will be useful to express symmetries (\ref{eq2_4}) as operators. 

We use a shorthand notation $u = (u_n)_{n \in \mathbb{Z}}$ for a bi-infinite sequence of shell velocities describing the full state of the system.
Let $u(t)$, $t \ge 0$, be an arbitrary function of time, denoted briefly as $u(\cdot)$.
Then, the space-scaling operator is defined for any $m \in \mathbb{Z}$ as
	\begin{equation}
	\mathcal{S}^m: u(\cdot) \mapsto \tilde{u}(\cdot),\quad 
	\tilde{u}_n(t) = \frac{u_{n+m}(t)}{\ell_m}.
	\label{eq2_S1}
	\end{equation}
Similarly, the time-scaling operator is defined for real parameters $\alpha > 0$ as
	\begin{equation}
	\mathcal{T}^\alpha: u(\cdot) \mapsto \tilde{u}(\cdot),\quad 
	\tilde{u}(t) = \frac{1}{\alpha} u\left(\frac{t}{\alpha}\right).
	\label{eq2_S2}
	\end{equation}
Finally, we define the time-shift (flow) operator for any $t \ge 0$ as
	\begin{equation}
	\Phi^{t}: u(\cdot) \mapsto \tilde{u}(\cdot), \quad \tilde{u}(s) = u(s+t).
	\label{eq2_S3b}
	\end{equation}
These operators are symmetries of the ideal system: if $u(t)$ solves Eqs.~(\ref{eq2_3}), then $\tilde{u}(t)$ also does. 
In particular, the scaling transformation (\ref{eq2_4}) is given by the composition 
	\begin{equation}
	\mathcal{S}_h =  \mathcal{T}^\alpha \circ \mathcal{S}, \quad \alpha = 2^{1-h}.
	\label{eq2_S3c}
	\end{equation}
Using Eqs.~(\ref{eq2_S1}) and (\ref{eq2_S2}) we express
	\begin{equation}
	\mathcal{S}_h^m: u(\cdot) \mapsto \tilde{u}(\cdot),\quad 
	\tilde{u}_n(t) = \ell_m^{-h} u_{n+m}(\ell_m^{1-h} t).
	\label{eq2_S1h}
	\end{equation}
One can verify the following composition and commutation relations 
	\begin{equation}
	\begin{array}{c}
	\displaystyle
	\mathcal{S}^{m_1} \circ \mathcal{S}^{m_2} = \mathcal{S}^{m_1+m_2}, \quad
	\mathcal{T}^{\alpha_1} \circ \mathcal{T}^{\alpha_2} = \mathcal{T}^{\alpha_1\alpha_2}, \quad
	\Phi^{t_1} \circ \Phi^{t_2} = \Phi^{t_1+t_2}, \\[3pt]
	\displaystyle
	\mathcal{S}^{m} \circ \mathcal{T}^\alpha = \mathcal{T}^\alpha \circ \mathcal{S}^{m}, \quad	
	\Phi^{t} \circ \mathcal{S}^{m} = \mathcal{S}^{m} \circ \Phi^{t}, \quad
	\Phi^{t} \circ \mathcal{T}^\alpha = \mathcal{T}^\alpha \circ \Phi^{t/\alpha}.
	\end{array}
	\label{eq2_S3}
	\end{equation}
Note that the last relation in Eq.~(\ref{eq2_S3}) shows that $\Phi^{t}$ and $\mathcal{T}^\alpha$ do not commute.

\subsection{Observables of the statistically stationary state} \label{subsec_obs}

Let us now introduce a general formalism that covers multi-time statistics in terms of observables and their temporal averages.
We consider observables as functionals $O: u(\cdot) \mapsto \mathbb{R}$ that associate a real number to a solution $u(t)$; the values can also be complex numbers or vectors. 
For each $O$ we define a respective family of observables as
	\begin{equation}
	O^{(h,m,t)} = O \circ \mathcal{S}_h^m \circ \Phi^t
	\label{eqSS_0}
	\end{equation}
for $h \in \mathbb{R}$, $m \in \mathbb{Z}$ and $t \ge 0$.
The last two operators are the space-time scaling (\ref{eq2_S1h}) and time shift (\ref{eq2_S3b}), and their composition acts as
	\begin{equation}
	\mathcal{S}_h^m \circ \Phi^t: u(\cdot) \mapsto \tilde{u}(\cdot),\quad 
	\tilde{u}_n(s) = \ell_m^{-h} u_{n+m}(t+\ell_m^{1-h} s).
	\label{eq2_S1hX}
	\end{equation}
Thus, if $O$ probes the solution at shells and times around the origin, then $O^{(h,m,t)}$ does so around the shell $m$ and time $t$.
Below are two illustrative examples:
	\begin{eqnarray}
	O[u(\cdot)] = |u_0(0)|^p & \Rightarrow &
	O^{(h,m,t)}[u(\cdot)] = \ell_m^{-ph}\big|u_m(t)\big|^p;
	\label{eqSS_0ex1} \\[3pt]
	O[u(\cdot)] = u_0(0)u_1(1) & \Rightarrow &
	O^{(h,m,t)}[u(\cdot)] = \ell_m^{-2h}\, u_m(t)
	u_{m+1}\big(t+\ell_m^{1-h}\big).
	\label{eqSS_0ex2}
	\end{eqnarray}

Let $u(t)$ be a given solution of the shell model. We define the time average of $O^{(h,m,t)}$ as
	\begin{equation}
	\big\langle O^{(h,m,t)} \big\rangle_t = \lim_{T \to \infty}\frac{1}{T}\int_0^T O^{(h,m,t)} [u(\cdot)] dt.
	\label{eqSS_1}
	\end{equation}
\newtext{For example, averaging the observables (\ref{eqSS_0ex1}) yields velocity moments of order $p$ with scale-dependent prefactors, while averages of the observables (\ref{eqSS_0ex2}) provide two-time and two-scale correlation functions.}
Numerical studies suggest that there exists an ergodic statistically stationary state~\cite{frisch1999turbulence}.
This means that the average values (\ref{eqSS_1}) do not depend on the choice of a (typical) initial condition and, therefore, on the choice of a solution $u(t)$. 

We say that $O^{(h,m,t)}$ is an inertial interval observable if it depends only on the data from the inertial interval. 
In examples (\ref{eqSS_0ex1}) and (\ref{eqSS_0ex2}), this means that the shell $m$ belongs to the inertial interval.
The Kolmogorov (K41) hypothesis proposed that the statistically stationary state is scale invariant in the inertial interval for the specific choice of the H\"older exponent $h = \sfrac{1}{3}$~\cite{kolmogorov1941local,frisch1999turbulence}. 
In our formalism this is the symmetry $\mathcal{S}_{\sfrac{1}{3}}$, and the K41 hypothesis states that
	\begin{equation}
	\big\langle O^{(\sfrac{1}{3},m,t)} \big\rangle_t \textrm{ do not depend on } m
	\label{eqSS_2}
	\end{equation}
for inertial interval observables.  
Combining (\ref{eqSS_0ex1}) and (\ref{eqSS_2}) yields the K41 scalings 
$\langle |u_m(t)|^p \rangle_t \propto \ell_m^{p/3}$. The latter contradicts to the well-established observation of anomalous scaling $\big\langle |u_n|^p \big\rangle_t \propto \ell_n^{\zeta_p}$, in which the exponents $\zeta_p$ depend nonlinearly on $p$. It follows that the K41 as well as any other scaling symmetry $\mathcal{S}_h$ is broken in the statistics of the inertial interval.

\section{Hidden scaling symmetry}
\label{sec_HS}

In this section, we introduce the concept of hidden symmetry following the results obtained in~\cite{mailybaev2021hidden,mailybaev2020hidden,mailybaev2022shell,mailybaev2023hidden}; see also Section~\ref{secI_HS}.
Extending the operator formalism of the previous section, we introduce a projector that maps the original solutions into rescaled ones. 
Next, we define the hidden symmetry operator for the rescaled dynamics. Finally, we express the hidden symmetry in terms of observables.

\subsection{Projector}

We introduce the operator $\mathcal{P}$, which defines a rescaled function $U(\tau)$, $\tau \ge 0$, as
    \begin{equation}
    \mathcal{P}: u(\cdot) \mapsto U(\cdot), \quad
    U(\tau) = \frac{u(t)}{A[u(t)]},\quad
    \tau = \int_0^t A[u(s)] \, ds.
    \label{eq3_2}
    \end{equation}
This operator combines the time-dependent state normalization with the implicit time transformation, where the scaling factor $A[u]$ is chosen in the form
    \begin{equation}
    A[u] = \sqrt{\sum_{j \ge 0} 8^{-j}|u_{-j}|^2}.
    \label{eq3_1}
    \end{equation}    
One can see that the rescaled function $U(\tau)$ satisfies the normalization condition
    \begin{equation}
    A[U(\tau)] = 1
    \label{eq3_1b}
    \end{equation}
at all times $\tau \ge 0$.
It follows that the subsequent projection $\mathcal{P}[U(\cdot)] = U(\cdot)$ acts as an identity, i.e. the operator $\mathcal{P}$ is a projection.
Note that the specific form (\ref{eq3_1}) of the scaling factor is taken for simplicity: it contains a positive series that converges geometrically~\cite{mailybaev2023hidden}. 
However, any similar (local, positive and homogeneous) expression for $A[u]$ can be used; see~\cite{mailybaev2020hidden} for the general approach. 

Next we introduce the operator acting on rescaled functions $U(\tau)$ as
	\begin{equation}
	\mathcal{H}: U(\cdot) \mapsto \tilde{U}(\cdot), \quad
	\tilde{U}_n(\tilde\tau) = \frac{U_{n+1}(\tau)}{\sqrt{|U_1(\tau)|^2+8^{-1}}}, \quad
	\tilde\tau = 2\int_0^{\tau} \sqrt{|U_1(s)|^2+8^{-1}}\, ds.
	\label{eq3_4}
	\end{equation}
Similarly to Eq.~(\ref{eq3_2}), this is the time-dependent state normalization with the implicit time transformation.
The operators $\mathcal{P}$ and $\mathcal{H}$ satisfy the commutation relations
	\begin{equation}
	\mathcal{P} \circ \mathcal{S} = \mathcal{H} \circ \mathcal{P}, \quad
	\mathcal{P} \circ \mathcal{T}^\alpha = \mathcal{P}.
	\label{eq3_4b}
	\end{equation}
The proof of these relations is elementary; see \cite{mailybaev2020hidden,mailybaev2023hidden}. 
Combining Eqs.~(\ref{eq2_S3c}) and (\ref{eq3_4b}) we have
	\begin{equation}
	\mathcal{P} \circ \mathcal{S}_h = \mathcal{H} \circ \mathcal{P}
	\label{eq3_4c}
	\end{equation}
for any $h \in \mathbb{R}$. Relation (\ref{eq3_4c}) means that all space-time scaling symmetries $\mathcal{S}_h$ are projected to the single operator $\mathcal{H}$. Namely, consider two functions related as $\mathcal{S}_h: u(\cdot) \mapsto \tilde{u}(\cdot)$. Then their projections $U(\cdot) = \mathcal{P}[u(\cdot)]$ and $\tilde{U}(\cdot) = \mathcal{P}[\tilde{u}(\cdot)]$ are related as $\mathcal{H}: U(\cdot) \mapsto \tilde{U}(\cdot)$.
This ``fusion'' of symmetries for all H\"older exponents is the central property of our projection. 

As an example, let us compute the functions
	\begin{equation}
	U^{(m)}(\cdot) = \mathcal{H}^m[U(\cdot)].
	\label{eq3_A1}
	\end{equation}
Using Eq.~(\ref{eq3_4b}), one derives
	\begin{equation}
	U^{(m)}(\cdot) = \mathcal{H}^m \circ \mathcal{P}[u(\cdot)] = \mathcal{P} \circ \mathcal{S}^m[u(\cdot)] = \mathcal{P}[u^{(m)}(\cdot)], \quad
	u^{(m)}(\cdot) = \mathcal{S}^m[u(\cdot)].
	\label{eq3_A1d}
	\end{equation}
Then the expressions (\ref{eq2_S1}) and (\ref{eq3_2}) yield the components of $U^{(m)}(\tau)$ in the form
	\begin{equation}
	U^{(m)}_N(\tau) = \frac{u_{N+m}(t)}{\ell_mA_m(t)}, \quad 
	\tau = \int_0^t A_m(s) ds, \quad
	A_m(t) = \frac{1}{\ell_m} \sqrt{\sum_{j \ge 0} 8^{-j}|u_{m-j}(t)|^2},
	\label{eq3_A2}
	\end{equation}
where we denoted $A_m(t) = A[u^{(m)}(t)]$.
Physical meaning of the function $U^{(m)}(\tau)$ follows from the estimate $A_m(t) \sim |u_m|/\ell_m$, which is an effective frequency of velocity fluctuations at shell $m$ and time $t$. Then $U^{(m)}(\tau)$ is the velocity sequence normalized with respect to its amplitude $\ell_m A_m(t) \sim |u_m|$ at shell $m$, while the rescaled time $\tau$ measures the integrated number of effective turnover times.

Finally, one can check that the projector (\ref{eq3_2}) composed with the time-shift operator (\ref{eq2_S3b}) yields
	\begin{equation}
	\mathcal{P} \circ \Phi^t[u(\cdot)] = \Phi^\tau \circ \mathcal{P}[u(\cdot)], \quad \tau = \int_0^t A[u(s)] \, ds.
	\label{eq3_4d}
	\end{equation}
Relations (\ref{eq3_4b}), (\ref{eq3_4c}) and (\ref{eq3_4d}) can be seen as kinematic properties of the rescaling procedure, because they refer to operators themselves and do not depend on the system dynamics.

\subsection{Hidden symmetry of a rescaled ideal system}

Now let us describe the dynamic properties of our projection. 
Assuming that components of $u(t)$ solve the ideal system (\ref{eq2_3}), one derives the rescaled ideal system (\ref{eq3_3}) for the corresponding rescaled velocities $U(\tau)$.
The derivation of Eq.~(\ref{eq3_3}) is elementary and uses relations~(\ref{eq3_2}) and (\ref{eq3_1}); we refer to~\cite{mailybaev2023hidden} for more details. The same rescaled system (\ref{eq3_3}) follows for $U^{(m)}(\tau)$ from Eq.~(\ref{eq3_A1}), because $U^{(m)}(\cdot)$ is a projection of another solution $u^{(m)}(\cdot) = \mathcal{S}^m[u(\cdot)]$. 
In fact, the operator $\mathcal{H}$ is a symmetry of the system (\ref{eq3_3}): if $U(\tau)$ solves Eq.~(\ref{eq3_3}), then $\mathcal{H}^m[U(\cdot)]$ yields another solution. 
One can verify this invariance by a direct calculation using relations (\ref{eq3_4}); see also~\cite{mailybaev2023hidden}. 
We call the operator $\mathcal{H}$ a hidden symmetry of the rescaled ideal system (\ref{eq3_3}). 

Now let us consider a solution $u(t)$ of the full shell model (\ref{eq2_1}). 
Then the ideal equations~(\ref{eq2_3}) are valid asymptotically for shells in the inertial interval. 
Let us extend the concept of inertial interval to the respective rescaled solutions $U^{(m)}(\tau)$ given by expressions (\ref{eq3_A2}).
We say that the component $U^{(m)}_N(\tau)$ belongs to the inertial interval if it is expressed in terms of shell velocities $u_n(t)$ from the inertial interval. 
It follows from Eq.~(\ref{eq3_A2}) that this requirement is satisfied if both shells $m$ and $N+m$ belong to the inertial interval; recall that the series in the last expression of Eq.~(\ref{eq3_A2}) converges geometrically and is thus determined by shells close to $m$.
Hence, the ideal rescaled system (\ref{eq3_3}) is valid asymptotically for the rescaled velocities $U^{(m)}_N(\tau)$ in the inertial interval. 

It was conjectured and confirmed numerically that the stationary statistics of the shell model restores the hidden symmetry in the inertial interval~\cite{mailybaev2021hidden}. 
In terms of rescaled velocities (\ref{eq3_A2}), this means that the statistics of inertial interval components $U^{(m)}_N$ does not depend on $m$.
Conceptually, the conjecture of restored hidden symmetry modifies the original Kolmogorov hypothesis by replacing the K41 symmetry $\mathcal{S}_{\sfrac{1}{3}}$ with the hidden symmetry $\mathcal{H}$.

Another important aspect of the hidden symmetry is its universality with respect to forcing and viscosity: the inertial interval statistics of the rescaled velocities $U^{(m)}_N$ remains the same if either the boundary (forcing) or viscous terms in the shell model change.
This universality was confirmed numerically in~\cite{mailybaev2023hidden}.

\subsection{Observables of the hidden symmetric state}

Let us formulate the hidden symmetry hypothesis in terms of observables. 
We consider a general observable as a functional $O_{\star}: U(\cdot) \mapsto \mathbb{R}$. For each $O_\star$ we introduce a family of observables as
	\begin{equation}
	O_\star^{(m,\tau)} = O_\star \circ \Phi^\tau \circ \mathcal{H}^m,
	\label{eqSS_0R}
	\end{equation}
where $m \in \mathbb{Z}$ and $\tau \ge 0$.
Here the action of the last two operators is expressed using Eq.~(\ref{eq3_A1}) as 
	\begin{equation}
	\Phi^\tau \circ \mathcal{H}^m: U(\cdot) \mapsto \tilde{U}(\cdot), \quad
	\tilde{U}(s) = U^{(m)}(\tau+s),
	\label{eqSS_0RX}
	\end{equation}
which combines the hidden-symmetry transformation with a time shift.
Given a rescaled solution $U(\tau)$, we define the time average of $O_\star^{(m,\tau)}$ as
	\begin{equation}
	\big\langle O_\star^{(m,\tau)} \big\rangle_\tau = \lim_{T \to \infty}\frac{1}{T}\int_0^T O_\star^{(m,\tau)} [U(\cdot)] d\tau.
	\label{eqSS_1R}
	\end{equation}
Just like in Section~\ref{subsec_obs}, the existence of an ergodic statistically stationary state implies that the mean values (\ref{eqSS_1R}) do not depend on the choice of a (typical) rescaled solution $U(\tau)$.

We say that the observable $O_\star^{(m,\tau)}$ expressed by Eqs.~(\ref{eqSS_0R}) and (\ref{eqSS_0RX}) belongs to the inertial interval if it depends only on the inertial interval components $U_N^{(m)}(\tau)$. 
Then the hidden symmetry hypothesis states that the averages of inertial interval observables
	\begin{equation}
	\big\langle O_\star^{(m,\tau)} \big\rangle_\tau \textrm{ do not depend on } m.
	\label{eqSS_2R}
	\end{equation} 
Moreover, the universality property of the hidden symmetric state implies that the mean values (\ref{eqSS_2R}) do not depend on a specific forcing mechanism and viscosity.

It is useful to give a more precise mathematical formulation for the hidden symmetry hypothesis (\ref{eqSS_2R}), replacing the asymptotic condition of the inertial interval with explicit limits. 
Recalling the condition (\ref{eq2_II}), the inertial interval can be represented by two consequitive limits. 
First, one takes the limit $\mathrm{Re} \to \infty$ in which the viscous scale $\eta \to 0$ vanishes. 
Then, one takes the limit $m \to \infty$, which yields the vanishing observation scale $\ell_m \to 0$.
Taking the limit $m \to \infty$ after $\mathrm{Re} \to \infty$ ensures that the scale $\ell_m$ belongs to the inertial interval. 
Therefore, the alternative form of the condition (\ref{eqSS_2R}) can be expressed as the double limit
	\begin{equation}
	\lim_{m \to \infty} \lim_{\mathrm{Re} \to \infty} \big\langle O_\star^{(m,\tau)} \big\rangle_\tau = O_\star^\infty.
	\label{eqSS_2RA}
	\end{equation}
The hidden symmetry hypothesis then states that the limiting value $O_\star^\infty$ exists and is universal for any $O_\star$ from a (properly defined) general class of observables.

\section{Self-similarity in the inertial interval}
\label{sec_4}

The hypothesis of restored hidden symmetry (\ref{eqSS_2R}) is formulated in terms of rescaled solutions $U^{(m)}(\tau)$; see Eqs.~(\ref{eqSS_0R}) and (\ref{eqSS_0RX}). The latter are related to the original solution $u(t)$ by a sophisticated transformation (\ref{eq3_A2}) of both the shell variables and time. 
What does the hidden symmetry hypothesis imply for the statistics of $u(t)$? 
\newtext{In this section, we present the main result of this paper formulated as a universal self-similarity rule (\ref{eqSS_3x}), which provides a general answer to this question.
Later in Section~\ref{sec_5} we show how known and new scaling laws follow from this universal rule.

First, we recall that the anomalous exponents $\zeta_p$ are intrinsic characteristics of the hidden symmetric statistics; see Section \ref{secI_HS}.
These exponents are defined as Perron--Frobenius eigenvalues, as established in~\cite{mailybaev2020hidden,mailybaev2022shell,mailybaev2023hidden}. 

Now let $O$ be a time-scale homogeneous observable of degree $p \in \mathbb{R}$, by which we mean that
	\begin{equation}
	O \circ \mathcal{T}^\alpha = \alpha^{-p} O
	\textrm{ \ for any \ } \alpha > 0. 
	\label{eqSS_3d}
	\end{equation}
We used a negative power in our definition (\ref{eqSS_3d}) to make $p$ match the orders of the structure functions; see Section~\ref{sec_5} below.
Assuming the hidden symmetry hypothesis, we state that
	\begin{equation}
	\big\langle O^{(\zeta_p/p,m,t)} \big\rangle_t = f_p \, C[O]
	\label{eqSS_3x}
	\end{equation}
in the inertial interval. This is our universal self-similarity rule. By self-similarity we mean that the averages do not depend on the observation scale $\ell_m$, which is achieved by associating the time-scale homogeneity degree $p$ to the H\"older exponent $h = \zeta_p/p$.
The quantity $C[O]$ on the right-hand side is the universal constant depending only on the observable $O$. 
Finally, there is a prefactor $f_p$ that depends on forcing and the degree of homogeneity $p$, but not on the observable itself.
In the case $p = 0$ we show that $f_0 = 1$ and Eq.~(\ref{eqSS_3x}) reduces to the form
	\begin{equation}
	\big\langle O^{(0,m,t)} \big\rangle_t = C[O],
	\label{eqSS_3xx}
	\end{equation}
where the undefined ratio $\zeta_p/p$ is replaced by zero.
The derivation of Eqs.~(\ref{eqSS_3x}) and (\ref{eqSS_3xx}) from the hidden scale invariance will be given in Section~\ref{sec_der}. 
}

The double limit representation (\ref{eqSS_2RA}) extends naturally to the relations (\ref{eqSS_3x}) and (\ref{eqSS_3xx}). 
Here the inertial interval condition is replaced by considering the limit $\displaystyle \lim_{m \to \infty} \lim_{\mathrm{Re} \to \infty}$ in the left-hand side.

\section{Applications}
\label{sec_5}

In this section we present several applications of the self-similarity relations (\ref{eqSS_3x}) and (\ref{eqSS_3xx}). 
We start with classical structure functions and then focus on multi-time observables.

\subsection{Single-time structure functions}
\label{subsec_5A}

Let us consider the observable $O[u(\cdot)] = |u_0(0)|^p$. Using the explicit form (\ref{eq2_S2}) of the  time-scaling operator, one can verify the time-scale homogeneity condition (\ref{eqSS_3d}).  Combining the self-similarity relation (\ref{eqSS_3x}) with the second expression in Eq.~(\ref{eqSS_0ex1}), we obtain 
	\begin{equation}
	S_p^{(m)} = \big\langle |u_m(t)|^p \big\rangle_t = f_p \, C[O] \, \ell_m^{\zeta_p}.
	\label{eqApp_1}
	\end{equation}
This is the well-known anomalous power law scaling in the inertial interval~\cite{frisch1999turbulence}. 

\begin{figure}[tp]
\centering
\includegraphics[width=1\textwidth]{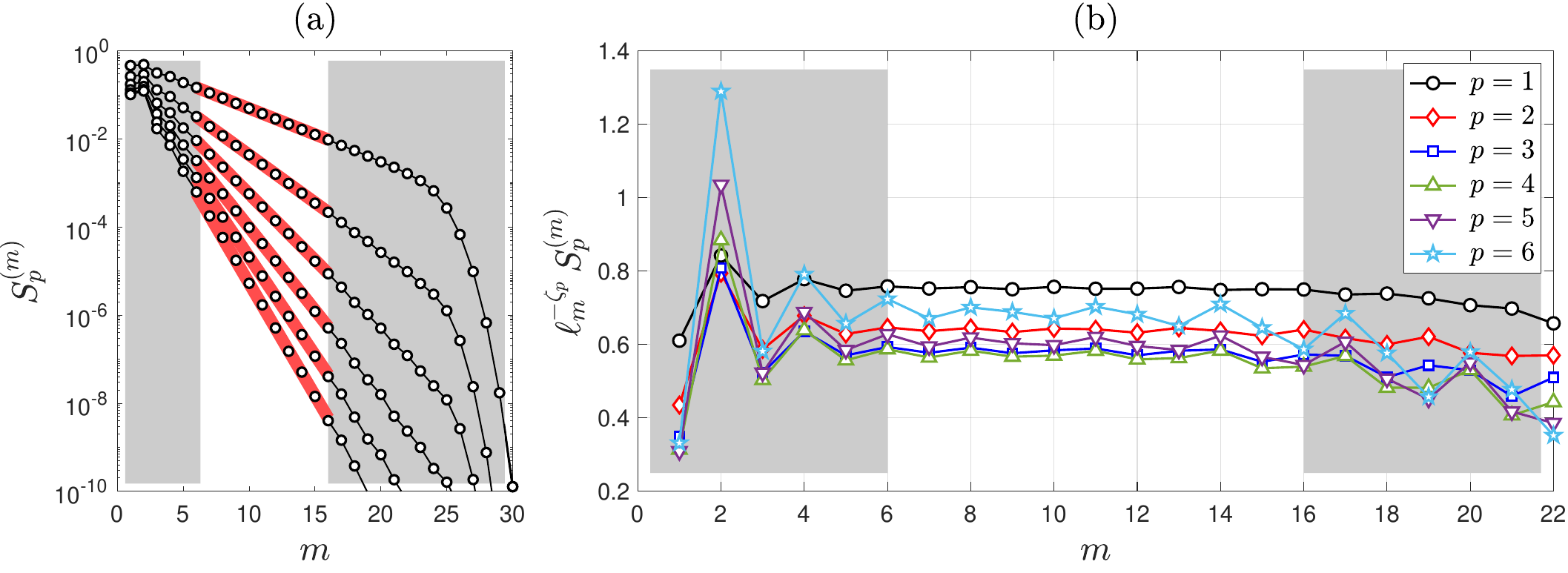}
\caption{(a) Empty dots are the values of the structure functions $S_p^{(m)}$ for $p = 1,\ldots,6$; note the vertical logarithmic scale. 
The thick red lines beneath the dots show the power laws (\ref{eqApp_1}). 
(b) Compensated structure functions $\ell_m^{-\zeta_p} S_p^{(m)}$ are asymptotically constant in the (white) inertial interval. In the gray regions, the hidden symmetry is broken by forcing (left) or viscosity (right).}
\label{fig1}
\end{figure}

Figure \ref{fig1}(a) shows the structure functions (\ref{eqApp_1}) obtained numerically for $p = 1,\ldots,6$. 
Our simulations are performed with $\mathrm{Re} = 10^{12}$ from K41 initial conditions with random phases. The statistics are calculated in the total time interval $T = 45000$, ignoring the initial transient dynamics at times $t  \le 100$.
System~(\ref{eq2_1}) is integrated for shells $n = 1,\ldots,35$ using the MATLAB solver $\mathtt{ode15s}$ with high accuracy. Our results are well resolved such that both numerical and statistical error bars (if plotted) would be smaller than graphical elements of the figures.

Figure~\ref{fig1}(b) verifies the self-similarity relations (\ref{eqApp_1}) by plotting the compensated values $\ell_m^{-\zeta_p} S_p^{(m)}$. 
For the anomalous exponents we use $\zeta_1 = 0.393$, $\zeta_2 = 0.72$, $\zeta_3 =  1$, $\zeta_4 =  1.25$, $\zeta_5 =  1.48$, $\zeta_5 =  1.695$
compatible with the earlier computations in~\cite{l1998improved,de2024extreme}. 
From Fig.~\ref{fig1}(b) we identify the inertial interval approximately as $8 \lesssim m \lesssim 20$,  
and highlight it by shading its outer part (forcing and dissipation ranges) in gray. 
In the inertial interval, the compensated values $\ell_m^{-\zeta_p} S_p^{(m)}$ demonstrate an asymptotic independence of $m$. 
Visible deviations from the constant asymptotics are finite-size effects: one can recognize deviations that arise in the forcing and dissipation regions and decay toward the center of the inertial interval.
Note that the dissipation range extends over a large range of shells $m \gtrsim 20$, larger than one might think from looking at the structure functions in Fig.~\ref{fig1}(a). 
In fact, viscous effects are strongly intermittent and their time-averaged influence can be small or large depending on the observable, for example, on the order of the structure function~\cite{frisch1993prediction}. We refer to \cite{mailybaev2023hidden} for the study of the dissipation range from the point of view of the hidden symmetry.

The power law (\ref{eqApp_1}) is naturally generalized to multi-scale observables.
For example, the observable $O[u(\cdot)] = |u_0(0)|^{p_1}|u_j(0)|^{p_2}$ of degree $p = p_1+p_2$ yields the scaling $\big\langle |u_m(t)|^{p_1}|u_{m+j} (t)|^{p_2} \big\rangle_t \propto \, \ell_m^{\zeta_p}$ for fixed $j$, $p_1$ and $p_2$. 
Also of interest may be large separations of the scales $\ell_m$ and $\ell_{m+j}$, which lead to so-called fusion rules~\cite{l1996towards,biferale2003shell} for structure functions. We do not study fusion rules here, but note that the Perron--Frobenius approach of Section~\ref{subsec_6C} potentially extends to this case.

\subsection{Multi-time structure functions}
\label{subsec_5B}

Equation (\ref{eqSS_0ex2}) gives an example of a two-time observable.
However, this observable is not time-scale homogeneous, like any other multi-time observable with constant time lags.

An example of multi-time observable, which is time-scale homogeneous of degree $p$, is
	\begin{equation}
	O_{2a}[u(\cdot)] = \big|u_0(\Delta)-u_0(0)\big|^p, \quad \Delta = \frac{1}{A[u(0)]},
	\label{eqAp2_1a}
	\end{equation}
as one can check using the condition (\ref{eqSS_3d}) and expression (\ref{eq3_1}).
The respective family of observables given by Eqs.~(\ref{eqSS_0}) and (\ref{eq2_S1hX}) express moments of two-time fluctuations of shell velocity $u_m(t)$ as
	\begin{equation}
	O_{2a}^{(h,m,t)}[u(\cdot)] = \ell_m^{-2h}\, \big|u_{m}\big(t+\Delta^{(m,t)}\big)-u_m(t)\big|^p,
	\label{eqAp2_1b}
	\end{equation}
where the time lag is expressed using $A_m(t) = A[u^{(m)}(t)]$ from Eq.~(\ref{eq3_A2}) in the form
	\begin{equation}
	\Delta^{(m,t)} 
	= \frac{1}{A_m(t)}
	= \frac{\ell_m}{\sqrt{\sum_{j \ge 0} 8^{-j}|u_{m-j}(t)|^2}}.
	\label{eqAp2_2f}
	\end{equation}
Note that the property of time-scale homogeneity is ensured by taking time differences equal to effective turnover times $\Delta^{(m,t)} \sim \ell_m/|u_m(t)|$. 
The important consequence is that the resulting time lag is not constant and depends on both the shell number and time.
\newtext{We note that some previous studies have already pointed out the need to use local turnover times in statistical analysis; see e.g.~\cite{benzi2004gibbs,domingues2024data}. }

\begin{figure}[tp]
\centering
\includegraphics[width=0.99\textwidth]{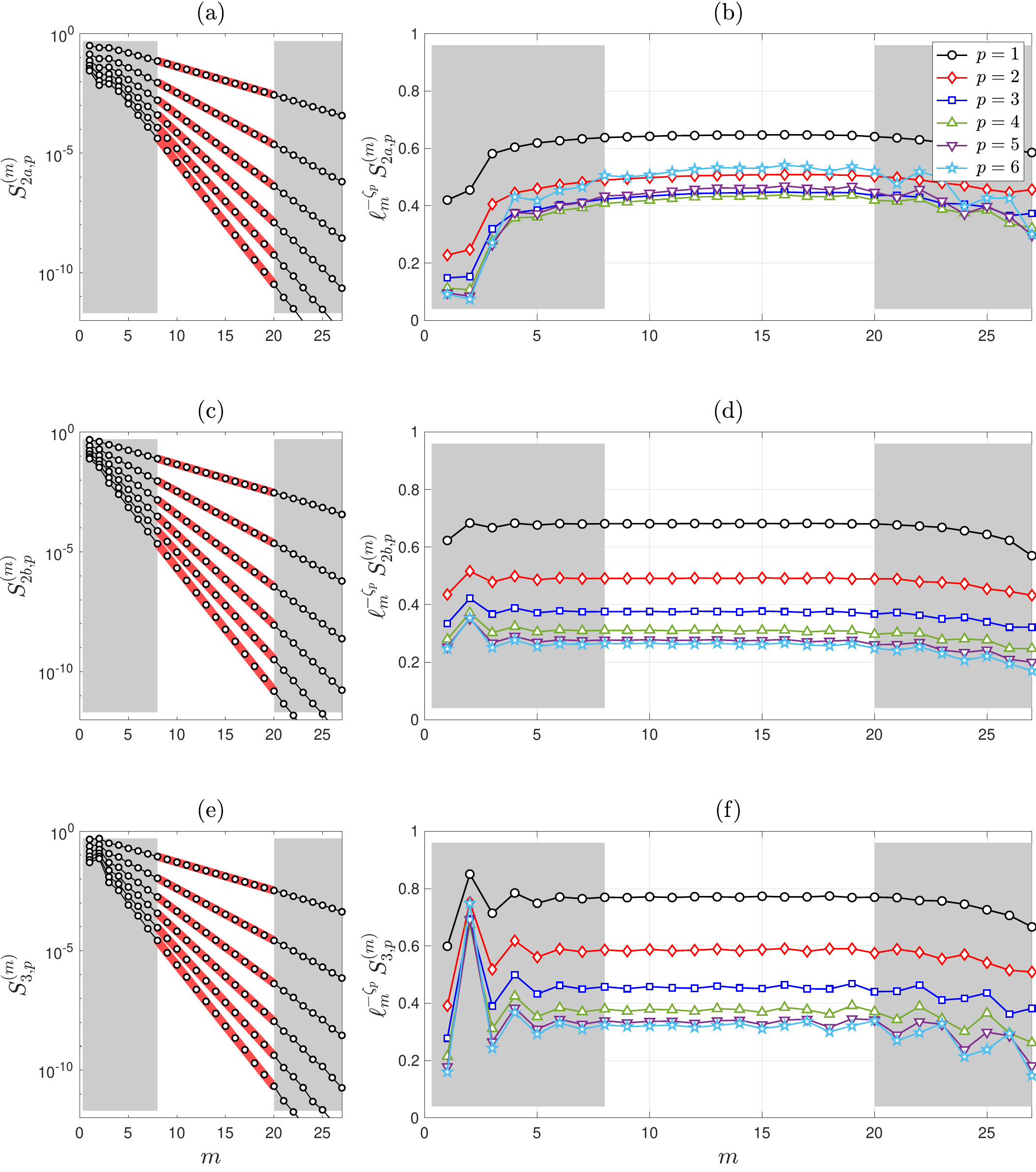}
\caption{Same plots as in Fig.~\ref{fig1} for the multi-time structure functions: (a,b) $S_{2a,p}^{(m)}$, (c,d) $S_{2b,p}^{(m)}$, (e,f) $S_{3,p}^{(m)}$. }
\label{fig2}
\end{figure}

The self-similarity relation (\ref{eqSS_3x}) applied to observables (\ref{eqAp2_1b}) yields the structure functions for two-time velocity fluctuations as
	\begin{equation}
	S_{2a,p}^{(m)} = \Big\langle \big| 
	u_{m}\big(t+\Delta^{(m,t)}\big)-u_m(t)\big|^p \Big\rangle_t = f_p \, C[O_{2a}] \, \ell_m^{\zeta_p}.
	\label{eqAp2_2b}
	\end{equation}
These structure functions have the same power law dependence as in Eq.~(\ref{eqApp_1}) but with a different coefficient $C[O_{2a}]$. 
We test this self-similarity relation by plotting the structure functions in Fig.~\ref{fig2}(a) together with the power laws $\propto \ell_m^{\zeta_p}$.
Figure~\ref{fig2}(b) presents the compensated values $\ell_m^{-\zeta_p} S_{2a,p}^{(m)}$ for $p = 1,\ldots,6$. As predicted, they demonstrate the asymptotic independence of $m$ in the inertial interval, up to decaying disturbances emanating from the forcing and dissipation regions. 

Another example of a time-scale homogeneous observable of degree $p$ is
	\begin{equation}
	O_{2b}[u(\cdot)] = |u_0(\Delta)|^{p/2} |u_1(0)|^{p/2}.
	\label{eqAp2_1}
	\end{equation}
The respective family of observables (\ref{eqSS_0}) takes the form
	\begin{equation}
	O_{2b}^{(h,m,t)}[u(\cdot)] = \ell_m^{-2h}\, |u_m(t)|^{p/2}
	\big|u_{m+1}\big(t+\Delta^{(m,t)}\big)\big|^{p/2}.
	\label{eqAp2_2}
	\end{equation}
The self-similarity relation (\ref{eqSS_3x}) yields the two-time two-scale correlation function 
	\begin{equation}
	S_{2b,p}^{(m)} = \Big\langle |u_m(t)|^{p/2}
	\big|u_{m+1}\big(t+\Delta^{(m,t)}\big)\big|^{p/2} \Big\rangle_t = f_p \, C[O_{2b}] \, \ell_m^{\zeta_p}.
	\label{eqAp2_2bb}
	\end{equation}
We test this self-similarity relation in Figs.~\ref{fig2}(c,d).

In the third example, we consider the three-time observable
	\begin{equation}
	O_3[u(\cdot)] = |u_0(0)|^{p/3} |u_0(\Delta)|^{p/3} |u_0(2\Delta)|^{p/3},
	\label{eqAp2_3}
	\end{equation}
which is time-scale homogeneous of degree $p$. The resulting self-similarity relation reads
	\begin{equation}
	S_{3,p}^{(m)} =  \Big\langle |u_m(t)|^{p/3}
	\big|u_{m}\big(t+\Delta^{(m,t)}\big)\big|^{p/3} 
	\big|u_{m}\big(t+2\Delta^{(m,t)}\big)\big|^{p/3} 
	\Big\rangle_t = f_p \, C[O_3] \, \ell_m^{\zeta_p}.
	\label{eqAp2_4}
	\end{equation}
This self-similarity relation is verified in Figs.~\ref{fig2}(e,f).

Other multi-time and multi-scale observables, which have the property of time-scale homogeneity, can be designed in a similar manner by changing the velocity terms and the time delays.
For example, one can use $\Delta = 1/|u(0)|$, which yields $\Delta^{(m,t)} = \ell_m/|u_m(t)|$. 

\newtext{

\subsection{Time-integrated correlation functions}
\label{subsecTI}

Another way to achieve the time-scale homogeneity is to use observables with integrated time differences.
For example, the observable 
	\begin{equation}
	O[u(\cdot)] = \int_0^{\Delta} \big| u_0(0)\big|^{p_1} \big|u_0(s)\big|^{p_2} s^{q-1} ds
	\label{eqAp2_B1}
	\end{equation}
is time-scale homogeneous of degree $p = p_1+p_2-q$. Then one derives
	\begin{equation}
	O^{(h,m,t)}[u(\cdot)] = \ell_m^{-q-hp}\int_0^{\Delta^{(m,t)}} \big| u_m(t)\big|^{p_1} \big|u_m(t+s)\big|^{p_2} s^{q-1} ds,
	\label{eqAp2_B1b}
	\end{equation}
and the respective self-similarity rule (\ref{eqSS_3x}) yields
	\begin{equation}
	\Big\langle \int_0^{\Delta^{(m,t)}} \big| u_m(t)\big|^{p_1} \big|u_m(t+s)\big|^{p_2} s^{q-1} ds \Big\rangle_t \propto \ell_m^{q+\zeta_p}.
	\label{eqAp2_B1c}
	\end{equation}

The upper limit of the integral in Eq.~(\ref{eqAp2_B1c}) can also be replaced by any multiple $M\Delta^{(m,t)}$ of the turnover time. 
Assuming that this average converges as $M \to \infty$, we express
 	\begin{equation}
	\lim_{T \to \infty} \Big\langle \int_0^T | u_m(t)|^{p_1} |u_m(t+s)|^{p_2} s^{q-1} ds \Big\rangle_t
	= \lim_{T \to \infty} \int_0^T \big\langle | u_m(t)|^{p_1} |u_m(t+s)|^{p_2}\big\rangle_t s^{q-1} ds 
	\propto \ell_m^{q+\zeta_p}.
	\label{eqAp2_B1SLx}
	\end{equation}
where we replaced the limit $M\Delta^{(m,t)} \to \infty$ by a constant time $T \to \infty$ and then commuted the $t$-averaging with the $s$-integration.
The resulting scaling law reads 
 	\begin{equation}
	\int_0^\infty \big\langle | u_m(t)|^{p_1} |u_m(t+s)|^{p_2} \big\rangle_t s^{q-1} ds 
	\propto \ell_m^{q+\zeta_p}.
	\label{eqAp2_B1SL}
	\end{equation}
Note that we just derived Eq.~(\ref{eqI_MS3}) from Section~\ref{subsec_MTS}. It is straightforward to extend this derivation from two-time to general multi-time correlators. For example, 
	\begin{equation}
	\int_0^\infty \cdots \int_0^\infty \big\langle | u_m(t)|^{p_0} |u_m(t+s_1)|^{p_1} \cdots |u_m(t+s_r)|^{p_r} \big\rangle_t 
	\prod_{i = 1}^r s_i^{q_i-1} ds_i \propto \ell_m^{q+\zeta_p},
	\label{eqAp2_B1SLG}
	\end{equation}
where $p = p_0+p_1+\cdots+p_r-q$ and $q = q_1+\cdots+q_r$.
These and similar scaling laws were derived in several previous works~\cite{l1997temporal,biferale1999multi,mitra2004varieties,pandit2008dynamic,ray2008universality} using the phenomenological multifractal model. 
The contribution of our self-similarity relation (\ref{eqSS_3x}) is two-fold: ($i$) it unifies these scaling laws through their relation to time-scale homogeneous observables and ($ii$) it provides their derivation from the hidden scaling symmetry of equations of motion.

}

\subsection{Single-time Kolmogorov multipliers}
\label{subsec_5C}

In this subsection we consider the self-similarity relation (\ref{eqSS_3xx}) corresponding to the special case $p = 0$. 
Using this relation we explain a universal self-similar statistics of Kolmogorov multipliers. Such properties of Kolmogorov multipliers were detected and analyzed numerically in several previous works \cite{benzi1993intermittency,eyink2003gibbsian,biferale2017optimal,PhysRevX.11.021063}, and their connection with hidden symmetry was revealed in \cite{mailybaev2022shell,mailybaev2022hidden,mailybaev2023hidden}.

The Kolmogorov multiplier at shell $m$ and time $t$ is defines as 
	\begin{equation}
	w_m(t) = \left|\frac{u_{m}(t)}{u_{m-1}(t)}\right|. 
	\label{eqAp3_1KM}
	\end{equation}
Let us define the observables
	\begin{equation}
	O[u(\cdot)] = f\left(\left|\frac{u_{0}(0)}{u_{-1}(0)}\right|\right) \quad \Rightarrow \quad
	O^{(0,m,t)}[u(\cdot)] = f\left(\left|\frac{u_{m}(t)}{u_{m-1}(t)}\right|\right) = f\big(w_m(t)\big).
	\label{eqAp3_1}
	\end{equation}
One can see that $O$ is time-scale homogeneous of degree $0$ for any function $f:\mathbb{R} \mapsto \mathbb{R}$.
The cumulative distribution function of the multiplier $F_m(x) = \mathrm{P}(w_m \le x)$ is obtained as the average 
	\begin{equation}
	F_m(x) = \big\langle O^{(0,m,t)} \big\rangle_t, 
	\label{eqAp3_2}
	\end{equation}
where the observable is given by Eq.~(\ref{eqAp3_1}) with the indicator function $f = \mathbf{1}_{[0,x]}$. 
Using the self-similarity relation (\ref{eqSS_3xx}), we conclude that the multipliers have the same probability distribution at all shells of the inertial interval, 
and this distribution is universal with respect to forcing and viscosity. 
This conclusion naturally extends to a mutual distributions of several multipliers.

Similar argument applies to the phases multipliers defined as \cite{benzi1993intermittency,eyink2003gibbsian}
	\begin{equation}
	\varphi_m(t) = \arg \big[ u_{m-1}(t)u_{m}(t)u_{m+1}^*(t)\big]. 
	\label{eqAp3_3def}
	\end{equation}
In this case we choose the observables
	\begin{equation}
	O[u(\cdot)] = f\big(\arg \left[ u_{-1}(0) u_0(0) u_1^*(0) \right] \big) \quad \Rightarrow \quad
	O^{(0,m,t)}[u(\cdot)] = f\big(\varphi_m(t)\big).
	\label{eqAp3_3}
	\end{equation}
 Again, one can see that  $O$ is time-scale homogeneous of degree $0$. 
The cumulative distribution function of the phase multiplier is obtained by the same average (\ref{eqAp3_2}) applied to the observables (\ref{eqAp3_3}) with the indicator function $f = \mathbf{1}_{[-\pi,x]}$. The self-similarity rule (\ref{eqSS_3xx}) states that the probability distributions of phase multipliers are scale-independent in the inertial interval. 
 
\subsection{Multi-time Kolmogorov multipliers}
\label{subsec_5D}

Let us now extend the results of the previous subsection to multi-time multipliers. Using the turnover times $\Delta$ and $\Delta^{(m,t)}$ from Eqs.~(\ref{eqAp2_1a}) and (\ref{eqAp2_2f}), we define 
	\begin{equation}
	O[u(\cdot)] = f\left(\frac{u_0(\Delta)}{u_{0}(0)}\right) 
	\quad \Rightarrow \quad
	O^{(0,m,t)}[u(\cdot)] = f\left(\frac{u_m\big(t+\Delta^{(m,t)}\big)}{u_{m}(t)}\right).
	\label{eqAp4_1}
	\end{equation}
The observable $O^{(0,m,t)}$ corresponds to the two-time complex multiplier 
	\begin{equation}
	\hat{w}_m(t) =  \frac{u_m\big(t+\Delta^{(m,t)}\big)}{u_{m}(t)},
	\label{eqAp4_2b}
	\end{equation}
which measures the relative change of the shell variable $u_m(t)$ over one turnover time. 
The probability distribution of this multiplier can be expressed as
	\begin{equation}
	\mathrm{P}_m(\hat{w}_m \in A) = \big\langle O^{(0,m,t)} \big\rangle_t,
	\label{eqAp4_3}
	\end{equation}
where $A \subset \mathbb{C}$ is a measurable set and $f = \mathbf{1}_{A}$ is the corresponding indicator function.
Then the self-similarity relation (\ref{eqSS_3xx}) states that this probability distributions does not depend on scale in the inertial interval. 

\begin{figure}[tp]
\centering
\includegraphics[width=0.8\textwidth]{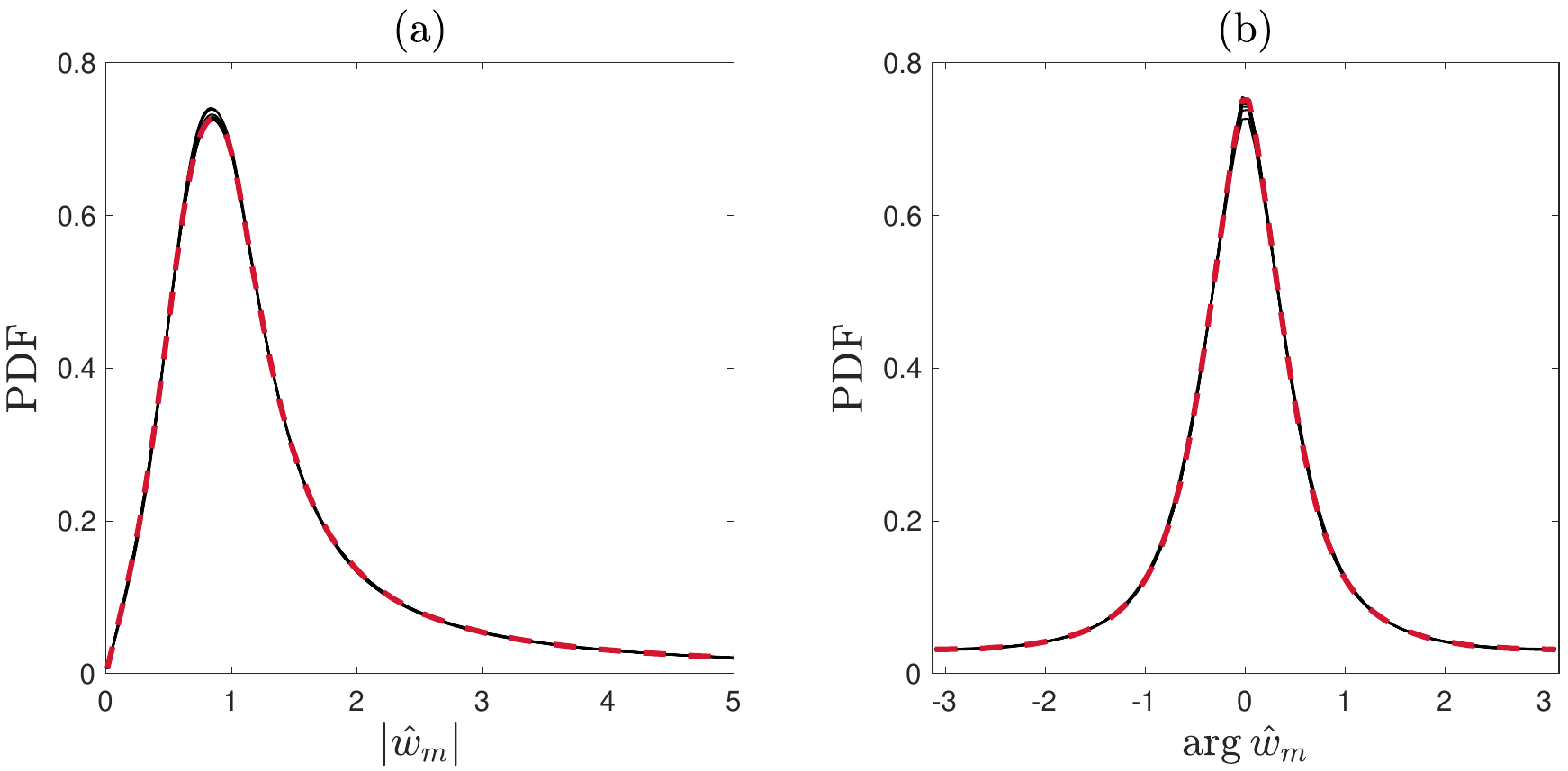}
\caption{Solid black lines are PDFs for (a) absolute values and (b) phases of two-time complex multipliers $\hat{w}_m(t)$. Each panel shows eleven (almost undistinguishable) PDFs for $m = 8,\ldots,18$. 
Dashed red lines show the same PDFs for $m = 11$ and different boundary conditions (\ref{eq2_BCb}).}
\label{fig3}
\end{figure}

Using numerical simulations, we plot in Fig.~\ref{fig3} the probability density functions (PDFs) for the absolute values (left panel) and the phases (right panel) of the multiples (\ref{eqAp4_2b}). These graphs confirm that the PDFs indeed collapse for the shells $m = 8,\ldots,18$ from the inertial interval.

\subsection{Universality with respect to viscosity and forcing}
\label{sec_5E}

\begin{figure}[tp]
\centering
\includegraphics[width=0.8\textwidth]{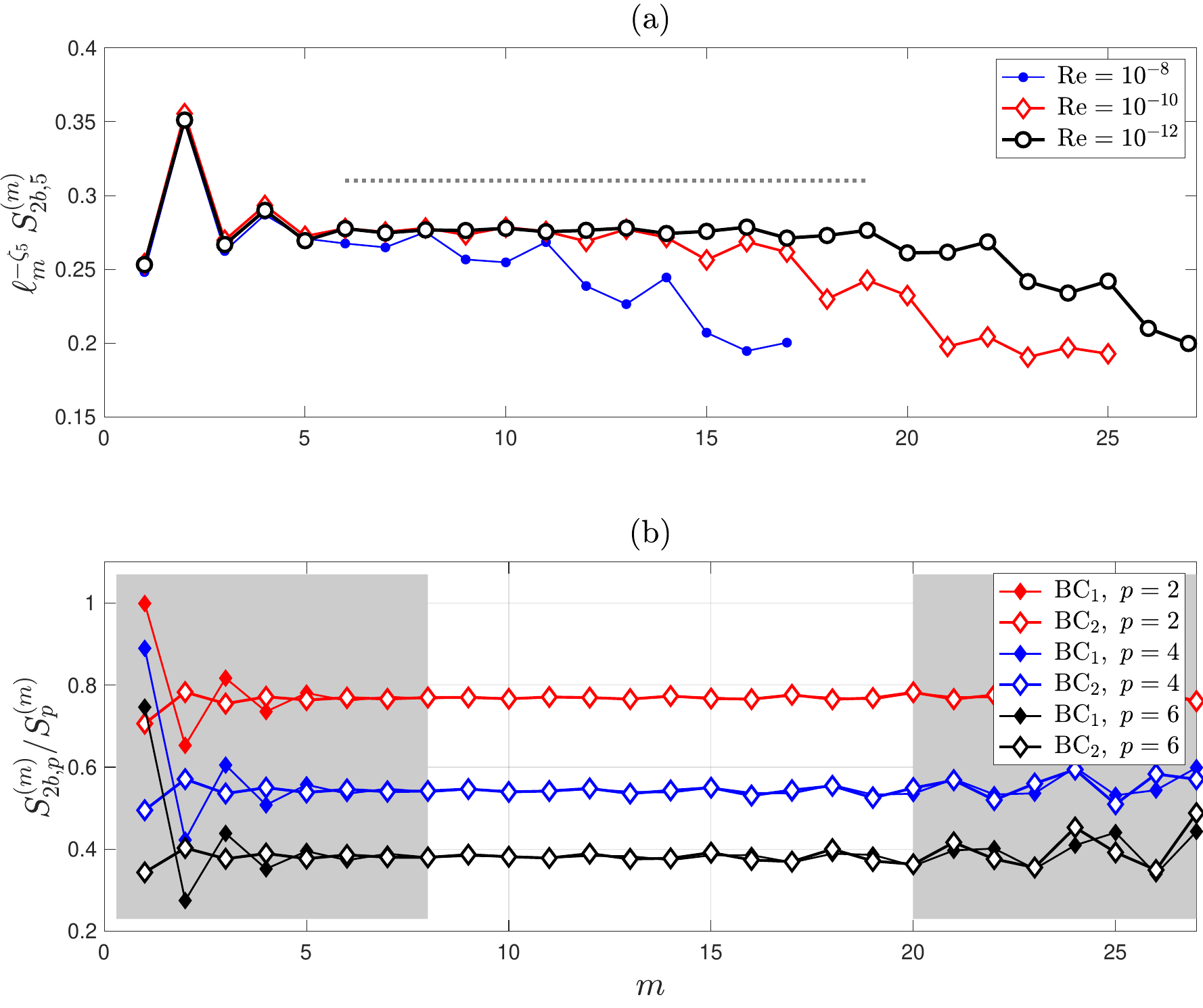}
\caption{(a) Compensated values $\ell_m^{-\zeta_p} S_{2b,p}^{(m)}$ of the two-time structure function (\ref{eqAp2_2bb}) of order $p = 5$. Numerical results with increasing Reynolds numbers demonstrate convergence to the same constant value in an increasing inertial interval. A dotted constant line is shown for visual reference. (b) Ratios $S_{2b,p}^{(m)}/S_p^{(m)}$ of the structure functions for two different boundary conditions, $\mathrm{BC}_1$ and $\mathrm{BC}_2$. Their values converge to the same constant in the inertial interval for every $p = 2, 4$ and $6$. }
\label{fig4}
\end{figure}

It is generally assumed that the statistically stationary state is independent of the Reynolds number outside the dissipation range, i.e., at scales of the inertial interval and the forcing range. 
In particular, such independence for the exponents $\zeta_p$ and prefactors in anomalous scaling laws (\ref{eqApp_1}) was verified numerically in~\cite{l1998universal}. 
Analogous study for the rescaled variables $U_N^{(m)}(\tau)$ was carried out in~\cite{mailybaev2023hidden}.
For our self-similarity relation (\ref{eqSS_3x}), this implies that both prefactors $f_p$ and $C[O]$ are universal with respect to dissipation (viscosity). 
As an example, let us consider the two-time structure function $S_{2b,p}^{(m)}$ from Eq.~(\ref{eqAp2_2bb}). Figure~\ref{fig4}(a) shows the compensated values $\ell_m^{-\zeta_p} S_{2b,p}^{(m)}$ for $p = 5$ evaluated numerically with the increasing Reynolds numbers $\mathrm{Re} = 10^8$, $10^{10}$ and $10^{12}$. This graphs demonstrate the convergence to the same constant value in the increasing inertial interval. 

Unlike viscosity, the change of forcing (boundary) conditions affects the statistically stationary state globally. 
Universal properties with respect to forcing for single-time structure functions were studied in~\cite{l2003strong}, and here we extend these results to multi-time statistics. 
For numerical tests, we use the second boundary condition of the form
	\begin{equation}
	u_0(t) \equiv 1/2, \quad u_{-1}(t) \equiv 1, \quad u_n(t) \equiv 0, \quad n < -1. 
	\label{eq2_BCb}
	\end{equation}

The only forcing-dependent part of our self-similarity relation (\ref{eqSS_3x}) is the prefactor $f_p$. 
It depends on the degree of homogeneity $p$, but not on a specific form of the observable. 
The other coefficient $C[O]$ in Eq.~(\ref{eqSS_3x}) does not depend on the forcing, but depends on the observable.
We test these properties by measuring the ratio of expressions (\ref{eqSS_3x}) for two different observables of the same degree $p$. 
Then the forcing-dependent factor $f_p$ cancels out and hence the ratio becomes independent of the forcing.
We confirm this independence numerically in Fig.~\ref{fig4}(b) by plotting the ratios 
	\begin{equation}
	\frac{S_{2b,p}^{(m)}}{S_p^{(m)}} = \frac{C[O_{2b}]}{C[O]}
	\label{eq2_BCR}
	\end{equation}
of the two- and one-time structure functions from Eqs.~(\ref{eqAp2_2bb}) and (\ref{eqApp_1}). 
The figure confirms that the ratios (\ref{eq2_BCR}) coincide asymptotically in the inertial interval for two different boundary conditions: $\mathrm{BC}_1$ from Eq.~(\ref{eq2_BC}) and $\mathrm{BC}_2$ from (\ref{eq2_BCb}). 

\begin{figure}[tp]
\centering
\includegraphics[width=0.5\textwidth]{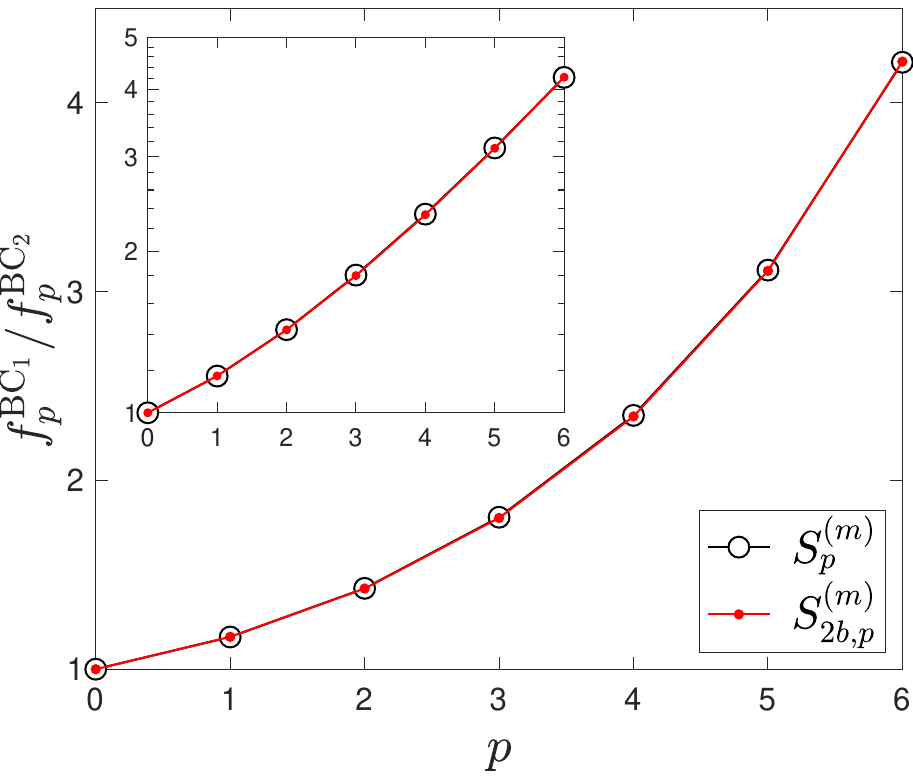}
\caption{Ratios of forcing-dependent prefactors $f_p^{\mathrm{BC}_1}/f_p^{\mathrm{BC}_2}$ for $p = 0,\ldots,6$ and fixed shell $m = 11$. 
The inset shows the same plots on a vertical logarithmic scale.}
\label{fig5}
\end{figure}

It is also interesting to see how the prefactor $f_p$ changes with the change of forcing (boundary conditions). We denote by $f_p^{\mathrm{BC}_1}$ and $f_p^{\mathrm{BC}_2}$ its values for the two boundary conditions under consideration. We find their ratio using the structure functions (\ref{eqApp_1}) and (\ref{eqAp2_2bb}) as
	\begin{equation}
	\frac{f_p^{\mathrm{BC}_1}}{f_p^{\mathrm{BC}_2}}
	= \frac{S_{p}^{(m)} \textrm{ for } \mathrm{BC}_1}{S_p^{(m)} \textrm{ for } \mathrm{BC}_2}
	= \frac{S_{2b,p}^{(m)} \textrm{ for } \mathrm{BC}_1}{S_{2b,p}^{(m)} \textrm{ for } \mathrm{BC}_2}.
	\label{eq2_BCf}
	\end{equation}
Here we specified that the structure functions in numerators and denominators are computed for $\mathrm{BC}_1$ and $\mathrm{BC}_2$, respectively.
The ratios (\ref{eq2_BCf}) obtained by numerical simulations for $m = 11$ are presented in Fig.~\ref{fig5}. 
They coincide for the one- and two-time structure functions, as predicted by Eq.~(\ref{eq2_BCf}). 
The inset shows the same plots on a vertical logarithmic scale, from which we conclude that the dependence of the ratios (\ref{eq2_BCf}) on $p$ is not a power law.
In particular, the dependence of $f_p$ on forcing cannot be reduced to a power of a single physical quantity (such as the energy flux~\cite{l2003strong}). 

Finally, for $p = 0$ the self-similarity rule (\ref{eqSS_3xx}) is universal with respect to the forcing mechanism. According to Sections~\ref{subsec_5C} and \ref{subsec_5D}, this implies that the statistics of multipliers does not depend on forcing. For one-time Kolmogorov multiples, this observation goes back to earlier works~\cite{benzi1993intermittency,eyink2003gibbsian}. Dashed red lines in Fig.~\ref{fig3} verify this universality for the two-time multiplier (\ref{eqAp4_2b}). 
 
\section{Derivation of the self-similarity relation}
\label{sec_der}

This section presents the derivation of the self-similarity relation (\ref{eqSS_3x}) and its special case (\ref{eqSS_3xx}) from the hidden scale invariance.
For convenience, we divide this derivation into four steps, considered in the following subsections. The first steps relates $\mathcal{P}$ to another projector.
The second steps express the observable $O^{(h,m,t)}$ in terms of a proper observable $O_\star^{(m,\tau)}$ for rescaled solutions. The third step does the same for their time averages. The derivation is completed in the fourth step, where we use the statistics of multipliers expressed in terms of Perron--Frobenius modes. 

Our derivation is based on formal assumptions about the developed turbulent state: existence of the inertial interval, the restored hidden symmetry, ergodicity and Perron--Frobenius asymptotics. 
Because of these assumptions, our derivation does not reach the level of a rigorous mathematical proof.
A possible direction towards a rigorous proof is to use the limit (\ref{eqSS_2RA}), which should be understood in a proper functional setting.

\subsection{Relationship between projectors}

Let us introduce the operator
	\begin{equation}
	\mathcal{R} : {u}(\cdot) \mapsto \hat{u}(\cdot), \quad
	\hat{u}(r) = \frac{1}{A[{u}(0)]} \, u\left( \frac{r}{A[{u}(0)]} \right).
	\label{eqPb_1f}
	\end{equation}
One can check using Eq.~(\ref{eq3_1}) that $A[\hat{u}(0)] = 1$ and hence $\mathcal{R}[\hat{u}(\cdot)] = \hat{u}(\cdot)$. This means that $\mathcal{R}$ is a projector. In this subsection we prove that, if $u(t)$ solves the ideal system (\ref{eq2_3}), then the projectors $\mathcal{R}$ and $\mathcal{P}$ are related as
	\begin{equation}
	\mathcal{R}  = \mathcal{Q} \circ \mathcal{P}.
	\label{eqPb_0c}
	\end{equation}
Here the transition operator $\mathcal{Q}$ acting on $U(\cdot) = \mathcal{P}[u(\cdot)]$ is defined as
	\begin{equation}
	\mathcal{Q}: U(\cdot) \mapsto \hat{u}(\cdot), \quad
	\hat{u}(r) = w(\tau)U(\tau), \quad
	r = \int_0^\tau \frac{ds}{w(s)}, \quad
	w(\tau) = \exp \bigg( \int_0^\tau R[U(s)] ds \bigg),
	\label{eqPa_2z}
	\end{equation}
with the function $R[U]$ from Eq.~(\ref{eq3_3}). The operator (\ref{eqPa_2z}) combines the time-dependent scaling of the state with the transformation of times $\tau \mapsto r$.

Let us proceed with the derivation. Let $U(\cdot) = \mathcal{P}[u(\cdot)]$. First, we derive the useful identify	
	\begin{equation}
	w(\tau) = \frac{A[u(t)]}{A[u(0)]},
    	\label{eqPb_n}
    	\end{equation}
where $\tau$ is given by Eq.~(\ref{eq3_2}).
Expressing $w(\tau)$ from  Eq.~(\ref{eqPa_2z}) and $R[U]$ from Eq.~(\ref{eq3_3}), we obtain
	\begin{equation}
	\begin{array}{rcl}
	\displaystyle
	\frac{d}{d \tau} \log w(\tau) 
	& = & \displaystyle
	\sum_{j \ge 0} 8^{-j} \Re \left( U^*_{-j}(\tau) B_{-j}[U(\tau)] \right)
	= \frac{1}{A^3[u(t)]}\sum_{j \ge 0} 8^{-j} \Re \left( u^*_{-j}(t) B_{-j}[u(t)] \right)
	\\[17pt]
	& = & \displaystyle
	\frac{1}{A^2[u(t)]} \frac{d}{d t} A[u(t)]  
	= \frac{1}{A[u(t)]} \frac{d}{d t} \log \frac{A[u(t)]}{A[u(0)]}.
	\end{array}
	\label{eqPb_4}
	\end{equation}
In this derivation, the second equality used expressions~(\ref{eq3_2}) with the quadratic form of the nonlinear term (\ref{eq2_3}), while the third equality follows from the expression (\ref{eq3_1}) and the ideal system (\ref{eq2_3}). Integrating the left-hand side of Eq.~(\ref{eqPb_4}) with respect to $\tau$ yields $\log w(\tau)$. Expressing $d\tau = A[u(t)] \, dt$ from Eq.~(\ref{eq3_2}), the same integration in the right-hand side of Eq.~(\ref{eqPb_4}) yields $\log \big( A[u(t)]/A[u(0)] \big)$. This proves the identity (\ref{eqPb_n}).

Let us now consider the operator $\mathcal{R}$. Using Eq.~(\ref{eqPb_n}) with relations~(\ref{eq3_2}), we express the function $\hat{u}(r)$ from Eq.~(\ref{eqPb_1f}) for $r = A[u(0)] t$ as
	\begin{equation}
	\hat{u}(r) = \frac{A[u(t)]}{A[u(0)]}\frac{u(t)}{A[u(t)]} = w(\tau) U(\tau), \quad
	\tau = \int_0^{t} A[u(s)] \, ds = \int_0^{r/A[u(0)]} A[u(s)] \, ds.
	\label{eqPb_R1}
	\end{equation}
Differentiating the second relation with the use of $t = r/A[u(0)]$ and Eq.~(\ref{eqPb_n}) yields 
	\begin{equation}
	\frac{d\tau}{dr} = \frac{A[u(r/A[u(0)])]}{A[u(0)]} = \frac{A[u(t)]}{A[u(0)]} = w(\tau). 
	\label{eqPb_R1tau}
	\end{equation}
This equality is integrated as 
	\begin{equation}
	r = \int_0^\tau \frac{ds}{w(s)}.
	\label{eqPb_R1tau2}
	\end{equation}
Combining Eqs.~(\ref{eqPb_R1}) and (\ref{eqPb_R1tau2}) with the definitions (\ref{eqPb_1f}) and (\ref{eqPa_2z}), we recover Eq.~(\ref{eqPb_0c}) as
	\begin{equation}
	\mathcal{R}[u(\cdot)] = \hat{u}(\cdot) = \mathcal{Q}[U(\cdot)] = \mathcal{Q} \circ \mathcal{P}[u(\cdot)].
	\label{eqPb_R1tau3}
	\end{equation}

\subsection{Reduction to a rescaled-velocity observable}

Let now $u(t)$ be a solution of the viscous shell model (\ref{eq2_1}), $U(\cdot) = \mathcal{P}[u(\cdot)]$ be the corresponding rescaled solution, and $O$ be a time-scale homogeneous observable of degree $p$. 
In this subsection we prove the relation
	\begin{equation}
	A_m^{-p}(t) \, O^{(h,m,t)}[u(\cdot)]
	= \ell_m^{(1-h)p} \, O_\star^{(m,\tau)} [U(\cdot)]
	\label{eqPa_2}
	\end{equation}
for any inertial interval observable $O^{(h,m,t)}$. Here $A_m(t)$ and $\tau$ are given by Eq.~(\ref{eq3_A2}) and 
	\begin{equation}
	O_\star = O \circ \mathcal{Q}
	\label{eqPa_2O}
	\end{equation}
with the operator $\mathcal{Q}$ defined by Eq.~(\ref{eqPa_2z}).

For the derivation we proceed as follows. Using Eqs.~(\ref{eqSS_0}) and (\ref{eq2_S3c}) with property (\ref{eqSS_3d}) we write
	\begin{equation}
	O^{(h,m,t)} = O \circ \mathcal{S}_h^m \circ \Phi^t
	= O \circ \mathcal{T}^{\ell_m^{h-1}} \circ \mathcal{S}^m \circ \Phi^t
	= \ell_m ^{(1-h)p} \, O \circ \mathcal{S}^m \circ \Phi^t.
	\label{eqPb_0a}
	\end{equation}
Since $\mathcal{T}^{\alpha} \circ \mathcal{T}^{1/\alpha}$ is the identity for any $\alpha > 0$, we similarly express
	\begin{equation}
	O 
	= O \circ  \mathcal{T}^{\alpha} \circ \mathcal{T}^{1/\alpha}
	= \alpha^{-p} O \circ \mathcal{T}^{1/\alpha}.
	\label{eqPb_1}
	\end{equation}
Combining Eqs.~(\ref{eqPb_0a}) and (\ref{eqPb_1}) and taking $\alpha = 1/A_m(t)$, we obtain
	\begin{equation}
	O^{(h,m,t)}[u(\cdot)]
	= \ell_m ^{(1-h)p} A_m^p(t)\, O \circ \mathcal{T}^{A_m(t)} \circ \mathcal{S}^m \circ\Phi^t [u(\cdot)].
	\label{eqPb_1g}
	\end{equation}
Using Eqs.~(\ref{eq2_S1}) and (\ref{eq2_S3b}) we write
	\begin{equation}
	\tilde{u}(\cdot) = \mathcal{S}^m \circ \Phi^t[u(\cdot)], \quad
	\tilde{u}_n(s) =  \frac{u_{m+n}(s+t)}{\ell_m}.
	\label{eq2_S1hPr}
	\end{equation}
Then, using Eqs.~(\ref{eq3_1}) and (\ref{eq3_A2}) we express 
	\begin{equation}
	A_m(t) = A[\tilde{u}(0)].
	\label{eqPb_1gg}
	\end{equation}
This relation combined with the definition (\ref{eq2_S2}) yield
	\begin{equation}
	\mathcal{T}^{A_m(t)} [\tilde{u}(\cdot)] = \mathcal{R}[\tilde{u}(\cdot)],
	\label{eqPb_1k}
	\end{equation}
where the operator $\mathcal{R}$ is defined in Eq.~(\ref{eqPb_1f}).

Since our relations refer to inertial interval observables, we can assume the ideal system (\ref{eq2_3}) to be valid in all calculations. 
Hence, we can use the identity (\ref{eqPb_0c}).
Combining Eqs.~(\ref{eqPb_1g}), (\ref{eq2_S1hPr}), (\ref{eqPb_1k}) with (\ref{eqPb_0c}), yields
	\begin{equation}
	\begin{array}{rcl}
	A_m^{-p}(t)\, O^{(h,m,t)}[u(\cdot)]
	& = & 
	\ell_m ^{(1-h)p} \, O \circ \mathcal{R} [\tilde{u}(\cdot)] 
	= \ell_m ^{(1-h)p} \, O \circ \mathcal{Q} \circ \mathcal{P} [\tilde{u}(\cdot)] 
	\\[5pt]
	& = & \ell_m ^{(1-h)p} \, O \circ \mathcal{Q} \circ \mathcal{P} \circ \mathcal{S}^m \circ\Phi^t[u(\cdot)] 
	= \ell_m ^{(1-h)p} \, O_\star \circ \mathcal{P} \circ \mathcal{S}^m \circ\Phi^t[u(\cdot)],
	\end{array}
	\label{eqPb_1mm}
	\end{equation}
where we substituted the definition (\ref{eqPa_2O}) in the last equality.
Using the commutation relations (\ref{eq2_S3}), (\ref{eq3_4b}) and (\ref{eq3_4d}), we have
	\begin{equation}
	\mathcal{P} \circ \mathcal{S}^m \circ\Phi^t[u(\cdot)] 
	= \mathcal{P} \circ\Phi^t \circ \mathcal{S}^m[u(\cdot)]
	= \Phi^\tau \circ \mathcal{P} \circ \mathcal{S}^m[u(\cdot)]
	= \Phi^\tau \circ \mathcal{H}^m \circ \mathcal{P}[u(\cdot)]
	= \Phi^\tau \circ \mathcal{H}^m[U(\cdot)].
	\label{eqPb_1m}
	\end{equation}
Substituting this relation into Eq.~(\ref{eqPb_1mm}) and recalling the definition (\ref{eqSS_0R}) yields Eq.~(\ref{eqPa_2}). 

\subsection{Relation between mean values}

In this subsection we prove the relation
	\begin{equation}
	\big\langle O^{(h,m,t)} \big\rangle_t 
	= \frac{\ell_m^{-ph}}{c_m} \Big\langle \Big(\prod_{N = -m+1}^0  X_{N\star}^{(m,\tau)}\Big)^{p-1} O_\star^{(m,\tau)} \Big\rangle_\tau, \quad
	c_m = \Big\langle \Big(\prod_{N = -m+1}^0  X_{N\star}^{(m,\tau)} \Big)^{-1} \Big\rangle_\tau,
	\label{eqPDa_0}
	\end{equation}
where the observables $X_{N\star}: U(\cdot) \mapsto \mathbb{R}$ are defined as
	\begin{equation}
	X_{N\star}[U(\cdot)] = \sqrt{8^{-1}+\frac{8^{-1}|U_N[0]|^2}{\sum_{j \ge 1} 8^{-j}|U_{N-j}(0)|^2}}.
	\label{eqPDa_2}
	\end{equation}

For the derivation, let us first show that
	\begin{equation}
	\frac{A_{m+N}(t)}{A_{m+N-1}(t)} = 2 X_{N\star}^{(m,\tau)}[U(\cdot)], \quad \tau = \int_0^t A_m(s) ds.
	\label{eqPDa_1}
	\end{equation}
These expressions describe a specific type of multipliers defined as ratios of the amplitudes $A_m(t)$~\cite{mailybaev2020hidden}.
By definition (\ref{eqSS_0R}), we have $X_{N\star}^{(m,\tau)} = X_{N\star} \circ \Phi^\tau \circ \mathcal{H}^m$. Using Eqs.~(\ref{eqSS_0RX}), (\ref{eqPDa_2}) and (\ref{eq3_A2}), elementary maniplations yield Eq.~(\ref{eqPDa_1}) as
	\begin{equation}
	\begin{array}{rcl}
	X_{N\star}^{(m,\tau)}[U(\cdot)] 
	& = & \displaystyle
	\sqrt{8^{-1}+\frac{8^{-1}|U_N^{(m)}[\tau]|^2}{\sum_{j \ge 1} 8^{-j}|U_{N-j}^{(m)}(\tau)|^2}}
	= \sqrt{\frac{8^{-1}\sum_{j \ge 0} 8^{-j}|U_{N-j}^{(m)}(\tau)|^2}{\sum_{j \ge 1} 8^{-j}|U_{N-j}^{(m)}(\tau)|^2}}
	\\[20pt]
	& = & \displaystyle
	\sqrt{\frac{\sum_{j \ge 0} 8^{-j}|U_{N-j}^{(m)}(\tau)|^2}{\sum_{j \ge 0} 8^{-j}|U_{N-1-j}^{(m)}(\tau)|^2}} 
	= \sqrt{\frac{\sum_{j \ge 0} 8^{-j}|u_{m+N-j}(t)|^2}{\sum_{j \ge 0} 8^{-j}|u_{m+N-1-j}(t)|^2}}
	\\[20pt]
	& = & \displaystyle
	 \frac{\ell_{m+N} A_{m+N}(t)}{\ell_{m+N-1} A_{m+N-1}(t)}
	= \frac{A_{m+N}(t)}{2 A_{m+N-1}(t)}.
	\end{array}
	\label{eqPDa_3}
	\end{equation}
Recall that $A_0(t) \equiv 1$ for the boundary conditions (\ref{eq2_BC}). Hence, using Eq.~(\ref{eqPDa_1}) we factorize 
	\begin{equation}
	A_m(t) = \prod_{N = -m+1}^0 \frac{A_{m+N}(t)}{A_{m+N-1}(t)}
	= \frac{1}{\ell_m} \prod_{N = -m+1}^0 X_{N\star}^{(m,\tau)}[U(\cdot)].
	\label{eqPF_1}
	\end{equation}
	
Using Eq.~(\ref{eqPF_1}) we express the observable $O^{(h,m,t)}$ from Eq.~(\ref{eqPa_2}) as 
	\begin{equation}
	O^{(h,m,t)}[u(\cdot)]
	= \ell_m^{-ph} \, \bigg(\prod_{N = -m+1}^0  X_{N\star}^{(m,\tau)}[U(\cdot)] \bigg)^p O_\star^{(m,\tau)} [U(\cdot)].
	\label{eqPa_2M}
	\end{equation}
Similarly, we write the relation between the times, $d\tau = A_m(t) dt$ from Eq.~(\ref{eq3_A2}), as
	\begin{equation}
	dt = \ell_m \bigg(\prod_{N = -m+1}^0  X_{N\star}^{(m,\tau)}[U(\cdot)] \bigg)^{-1} d\tau.
	\label{eqPa_2Mb}
	\end{equation}
Using Eqs.~(\ref{eqPa_2M}) and (\ref{eqPa_2Mb}), we express the average of $O^{(h,m,t)}$ in the form
	\begin{equation}
	\begin{array}{rcl}
	\big\langle O^{(h,m,t)} \big\rangle_t 
	& = & \displaystyle
	\lim_{T \to \infty}\frac{\int_0^T O^{(h,m,t)} [u(\cdot)] dt}{\int_0^T dt} 
	\\[20pt]
	& = & \displaystyle
	\ell_m^{-ph} \lim_{\tilde{T} \to \infty} \frac{
	\int_0^{\tilde{T}} 
	\bigg(\prod_{N = -m+1}^0  X_{N\star}^{(m,\tau)}[U(\cdot)] \bigg)^{p-1} O_\star^{(m,\tau)} [U(\cdot)] \, d\tau}{
	\int_0^{\tilde{T}} \bigg(\prod_{N = -m+1}^0  X_{N\star}^{(m,\tau)}[U(\cdot)] \bigg)^{-1} d\tau},
	\end{array}
	\label{eqPDa_10}
	\end{equation}
where the second equality is obtained by changing the integration time, $dt \mapsto d\tau$,  with the corresponding integration limit $\tilde{T} = \int_0^{T} A_m(s)ds \to \infty$ as $T \to \infty$. 
Dividing the numerator and denominator by $\tilde{T}$ and taking a separate limit $\tilde{T} \to \infty$ above and below, one obtains relation (\ref{eqPDa_0}).

\subsection{Integration using Perron--Frobenius modes}
\label{subsec_6C}

Let $x_N^{(m)}$ be a random variable whose distribution is determined by the stationary statistics of the observable $X_{N\star}^{(m,\tau)}[U(\cdot)]$. 
These observables are well-defined for $N \ge -m+1$. 
Without loss of generality we can set $x_N^{(m)} \equiv 1$ for $N \le -m$.
We denote a joint probability measure for these variables as $d\mu^{(m)}(x_\ominus)$, where $x_\ominus = \big(x_0,x_{-1},x_{-2},\ldots \big)$. 
Note that we have omitted the superscript $(m)$ for the arguments since it is present in the measure notation.
Next, we introduce the measures 
	\begin{equation}
	d\mu_p^{(m)}(x_\ominus ) = \frac{1}{c_m}\bigg( \prod_{N = -m+1}^0 x_N \bigg)^{p-1} d\mu^{(m)}(x_\ominus), \quad
	c_m = \int \bigg( \prod_{N = -m+1}^0 x_N \bigg)^{-1} d\mu^{(m)}(x_\ominus ).
	\label{eqPS3_1}
	\end{equation}
Using these measures, we write Eq.~(\ref{eqPDa_0}) as the integral
	\begin{equation}
	\big\langle O^{(h,m,t)} \big\rangle_t 
	= \ell_m^{-ph}
	\int \big\langle O_\star^{(m,\tau)}\big| x_\ominus \big\rangle_\tau d\mu_p^{(m)}(x_\ominus),
	\label{eqPS3_3}
	\end{equation}
where $\big\langle O_\star^{(m,\tau)}\big| x_\ominus \big\rangle_\tau$ denotes the conditional average of the observable $O_\star^{(m,\tau)}$. 

According to Eq.~(\ref{eqPDa_3}), the observable $X_{N\star}^{(m,\tau)}$ is determined by the rescaled velocity $U_{N}^{(m)}(\tau)$ and its close neighbors. 
Whenever $O_{\star}^{(m,\tau)}$ is the inertial interval observable, its conditional average $\big\langle O_\star^{(m,\tau)}\big| x_\ominus \big\rangle_\tau$ depends only  on the components of $x_\ominus$ from the inertial interval. In this case the hidden symmetry assumption (\ref{eqSS_2R}) extends to the conditional averages, i.e., $\big\langle O_\star^{(m,\tau)}\big| x_\ominus \big\rangle_\tau$ does not depend on $m$. Additionally, the universality of the hidden-symmetric state implies the universality of $\big\langle O_\star^{(m,\tau)}\big| x_\ominus \big\rangle_\tau$ with respect to forcing and dissipation.
Our derivation using Eq.~(\ref{eqPS3_3}) now relies on computing the measure $d\mu_p^{(m)}\big(x_\ominus\big)$ restricted to the inertial interval.
This measure describes the amplitude multipliers and therefore does not depend on the observable, except for its dependence on the degree of homogeneity $p$.

In earlier works~\cite{mailybaev2020hidden,mailybaev2022shell,mailybaev2023hidden} we used the hidden symmetry hypothesis and derived the asymptotic form of the measure (\ref{eqPS3_1}) in the inertial interval as
	\begin{equation}
	d\mu_p^{(m)}(x_\ominus) \approx f_p \, \lambda_p^m \, d\nu_p(x_\ominus).
	\label{eqPS3_2}
	\end{equation}
Here the quantity $\lambda_p > 0$ is the Perron--Frobenius eigenvalue and $d\nu_p(x_\ominus)$ is the corresponding eigenvector (probability measure), which are defined for specific positive operators in the framework of the statistically restored hidden symmetry. 
Both $\lambda_p$ and $d\nu_p(x_\ominus)$ are universal together with the hidden-symmetric state. 
The coefficient $f_p$ is not universal and depends on the forcing and degree of homogeneity $p$. 
In the case $p = 0$, Eq.~(\ref{eqPS3_1}) defines $d\mu_p^{(m)}(x_\ominus)$ as a probability measure. Then, Eq.~(\ref{eqPS3_2}) determines 
	\begin{equation}
	f_0 = \lambda_0 = 1.
	\label{eqPS3_4f}
	\end{equation}
We do not repeat these derivations here and refer the reader to~\cite[Sec. III D]{mailybaev2023hidden} for details.

Substituting the asymptotic form (\ref{eqPS3_2}) into Eq.~(\ref{eqPS3_3}) yields the self-similarity relation (\ref{eqSS_3x}) with 
	\begin{equation}
	\zeta_p = -\log_2 \lambda_p, \quad
	C[O] = \int \big\langle O_\star^{(m,\tau)}\big| x_\ominus \big\rangle_\tau d\nu_p(x_\ominus).
	\label{eqPS3_4}
	\end{equation}
The quantity $C[O]$ does not depend on $m$ and is universal due to the identical properties of both $\big\langle O_\star^{(m,\tau)}\big| x_\ominus \big\rangle_\tau$ and $d\nu_p(x_\ominus)$. Then the relation (\ref{eqSS_3xx}) for $p = 0$ follows from Eq.~(\ref{eqPS3_4f}). 

\newtext{Although $C[O]$ is usually a finite non-zero number, as shown in the examples of Section~\ref{sec_5}, there are two types of exceptions worth mentioning. 
The first exception concerns observables with $C[O] = 0$. This may happen, e.g., due to an additional symmetry. 
In this case, it is necessary to take into account the next order terms in the Perron--Frobenius approach, which provide larger scaling exponents.
Another exception refers to the case when $C[O] = \infty$, as defined by the divergent integral in Eq.~(\ref{eqPS3_4}). 
Such a situation may mean that the observable is not local and, therefore, is not an inertial interval observable.}

\section{Conclusion}
\label{sec_7}

\newtext{Equations of the ideal (inviscid) Sabra model of turbulence can be transformed by rescaling its velocities and time with respect to the instantaneous turnover time at a given scale. 
The resulting system acquires a new form of scaling symmetry called the hidden symmetry.
It was conjectured and confirmed numerically that the hidden symmetry is restored in the inertial interval of turbulent statistics~\cite{mailybaev2021hidden}. 
As a consequence, the anomalous scaling exponents of structure functions were derived as Perron--Frobenius eigenvalues~\cite{mailybaev2022shell,mailybaev2023hidden}.
The hidden symmetry of the Navier--Stokes turbulence was obtained in \cite{mailybaev2020hidden,mailybaev2022hidden}.

In this paper we derived a consequence of the restored hidden symmetry for the general (multi-time and multi-scale) statistics. 
It is formulated in terms of observables that are homogeneous with respect to time scaling. 
Any such observable is shown to be self-similar in the inertial interval with the H\"older exponent $h = \zeta_p/p$, where $p$ is the degree of homogeneity and $\zeta_p$ is the usual anomalous exponent. 
This universal self-similarity rule provides the unified formulation for a variety of scaling laws known from the phenomenological multifractal model. Note that our derivation is not phenomenological and relates the scaling laws to the exact symmetry of the ideal equations.
For this derivation, we developed and used an operator formalism for symmetries and observables.

As further applications, we derived scaling laws for multi-time structure functions and explained universal properties of multi-time Kolmogorov multipliers. 
Here the condition of time-scale homogeneity requires that time differences are measured in terms of instantaneous (time-dependent) turnover times.}
We construct examples of such observables and confirm their self-similarity by accurate numerical simulations. 

\vspace{2mm}\noindent\textbf{Acknowledgments.} 
This work was supported by CNPq grant 308721/2021-7, FAPERJ grant E-26/201.054/2022 and CAPES grant AMSUD3169225P. 

\bibliographystyle{plain}
\bibliography{refs}

\end{document}